\documentclass{article}
\usepackage{arxiv}
\usepackage[utf8]{inputenc} % allow utf-8 input
\usepackage[T1]{fontenc}    % use 8-bit T1 fonts
\usepackage{url}            % simple URL typesetting
\usepackage{booktabs}       % professional-quality tables
\usepackage{amsfonts}       % blackboard math symbols
\usepackage{nicefrac}       % compact symbols for 1/2, etc.
\usepackage{microtype}      % microtypography
\usepackage{graphicx}
\usepackage{natbib}
\usepackage{doi}
\usepackage{subfigure}
\usepackage{miller}
\usepackage{amsmath}
\usepackage{amssymb}
\usepackage{float}
\usepackage{color,soul}
\usepackage{mathtools}
\usepackage{appendix}
\usepackage{color}
\usepackage{tikz}
\usepackage{siunitx}
\usepackage{lineno}
\usepackage[normalem]{ulem}
\usepackage{upgreek}
\usepackage{siunitx}
\usepackage{fancyhdr}

\title{Establishing the relationship between generalized crystallographic texture and macroscopic yield surfaces using partial input convex neural networks}

\author{Lloyd van Wees \\
	Department of Mechanical Engineering\\
	The University of Alabama\\
	Tuscaloosa, AL 35487 \\
	\texttt{llvanwees@crimson.ua.edu.edu} \\
	\And
	Karthik Shankar \\
	Department of Mechanical Engineering\\
	The University of Alabama\\
	Tuscaloosa, AL 35487 \\
	\texttt{kshankar@crimson.ua.edu.edu} \\
        \And
	Jan N. Fuhg \\
	Department of Aerospace Engineering and Engineering Mechanics\\
	The University of Texas at Austin\\
	Austin, TX 78712 \\
	\texttt{jan.fuhg@utexas.edu} \\
        \And
	Nikolaos Bouklas \\
	Sibley School of Mechanical and Aerospace Engineering\\
	Cornell University\\
	Ithaca, NY 14853 \\
	\texttt{nb589@cornell.edu} \\
        \And
	Paul Shade \\
	Air Force Research Laboratory\\
	Materials and Manufacturing Directorate\\
	Wright Patterson AFB, OH 45433 \\
	\texttt{mark.obstalecki@us.af.mil} \\
        \And
	Mark Obstalecki \\
	Air Force Research Laboratory\\
	Materials and Manufacturing Directorate\\
	Wright Patterson AFB, OH 45433 \\
	\texttt{paul.shade.1@us.af.mil} \\
        \And
	Matthew Kasemer \\
	Department of Mechanical Engineering\\
	The University of Alabama\\
	Tuscaloosa, AL 35487 \\
	\texttt{mkasemer@eng.ua.edu} \\
}

\renewcommand{\shorttitle}{van Wees et. al. 2024}

\hypersetup{
pdftitle={Establishing the relationship between generalized crystallographic texture and macroscopic yield surfaces using partial input convex neural networks},
pdfsubject={q-bio.NC, q-bio.QM},
pdfauthor={Lloyd van Wees, Karthik Shankar, Jan N. Fuhg, Nikolaos Bouklas, Paul Shade, Mark Obstalecki, Matthew Kasemer},
pdfkeywords={Crystal plasticity, Macroscopic yield, Yield surfaces, Machine learning, Neural networks},
}

\begin{document}
\maketitle

\begin{abstract}
In this study, we present a methodology to predict the macroscopic yield surface of metals and metallic alloys with general crystallographic textures. In previous work, we have established the use of partially input convex neural networks (pICNN) as macroscopic yield functions of crystal plasticity simulations. However, this work was performed with an over-abundance of data, and on limited crystallographic textures. Here, we extend this study to approach more realistic material states (i.e., complex crystallographic textures), and consider data-availability as a major driver for our approach. We present our modified framework capable of handling generalized material states and demonstrate its effectiveness on samples with multi-modal textures deformed under plane stress conditions. We further describe an adaptive algorithm for the generation of training data as informed by the shape of yield surfaces to reduce the time for both the generation of training data as well as pICNN training. Finally, we will discuss errors in both training and test datasets, limitations, and future extensibility.
\end{abstract}

% keywords can be removed

\section{Introduction}
\label{sec:Introduction}

It is well established that the macroscopic behavior of materials is dependent on lower length-scale phenomena. Over the past 50 years, considerable work has been performed to establish relationships between the microstructure of metals and metallic alloys and their resulting macroscopic properties (such as yield strength or ductility). Perhaps most importantly, it has been found that the anisotropic single-crystal elastic-plastic behavior, coupled with the crystallographic texture (or the distribution of crystal orientations), has a profound impact on a sample's macroscopic properties~\citep{d1838introduction,Kocks2000}. The effect of texture is perhaps most evident in the degree to which it affects the anisotropic yield behavior of the material, best visualized in the yield surfaces of textured materials~\citep{backofen}. These aspects will ultimately dictate the behavior---and failure---of engineering components, and must be considered in design decisions.

However, there exist few widely-adopted models that consider how the microstructure effects the shape of the macroscopic yield surface, and indeed design engineers still widely utilize isotropic quadratic yield surfaces such as von Mises~\citep{mises} in their design considerations. Such models are often employed with generous factors of safety to---among other concerns---account for uncertainty with respect to the potential anisotropy present in the material. However, as modern design constraints necessitate the reduction of weight and material usage in engineering components to meet performance and sustainability demands, the need for more accurate yield prediction becomes more acute. Significant development towards the generation of models for anisotropic yield surfaces include the work of Hill~\citep{Hill1948,Bishop1951,Bishop1951b}, and later Barlat~\citep{Barlat1989,Barlat1991,Barlat2005} (as well as similar, derivative formulations~\citep{plunkett2008orthotropic,Liu2020}). However, these models do not directly consider the microstructure of the material and thus require careful (re)calibration depending on material and material state. It is thus difficult to implement these models in design considerations with confidence that they are broadly representative.

More recently, modeling frameworks have been developed that consider microstructure in various ways in an effort to understand and parameterize micromechanical behavior. The pinnacle of modern computational modeling of polycrystal deformation is crystal plasticity finite element modeling (CPFEM), which explicitly considers the microstructure of the material when predicting the microscopic and macroscopic mechanical response~\citep{Kasemer2017,Chatterjee2018,Kasemer2020,echlinonr}. Consequently, CPFEM simulations ultimately allow for the accurate prediction of anisotropic yield surfaces assuming a given/known microstructure. The issue, however, lies in CPFEM's relatively high computational cost, which hinders its employment in component-scale simulations. Consequently, it is of interest to up-scale the essential information of CPFEM simulations into data-driven constitutive models which may be incorporated in component-scale simulations. This approach would enable the efficient consideration of microstructural effects on yield at the component-scale.

Within the past decade, the use of machine learning algorithms to assimilate data in the fields of mechanics and materials science has accelerated dramatically. Regarding metals and metallic alloys, some studies have focused on the development of novel constitutive models at various material scales, including bridging the crystal and sample/component scales (i.e., structure-property relationships)~\citep{pagan,Meyer2022,Chung2024} as well as, specifically, the prediction of anisotropic yield surfaces~\citep{Hartmaier2020,vlassis2023geometric,Heidenreich2023,Heidenreich2023b,Nascimento2023,Shoghi2024,Shoghi2024b,Ghnatios2024,Jian2024}. In a previous study~\citep{fuhgcnn}, we focused on the development of convex yield functions by using partially input convex neural networks (pICNNs) trained with data from CPFEM simulations to relate the parameterized microstructural features of the material to the macroscopic response. This framework proved adept at relating crystallographic texture to the yield surface, though on problems with significantly reduced complexity to establish proof of function.

In this study, we extend the function of our pICNN framework to consider more complex, multi-dimensional material descriptions (i.e., generalized crystallographic texture) while at the same time reduce the requirements for data availability to the low-data regime, which translates to moderate computational costs for the generation of synthetic data. We will discuss the generation of a dataset of yield predictions via CPFEM simulations instantiated with samples comprised of complex, multi-modal crystallographic textures, including the strategic choice of a minimal number of simulations to accurately elucidate the yield surface via an adaptive data generation algorithm. We will further discuss extensions to the pICNN framework necessary to handle these more complex data. We will demonstrate how the pICNN framework is able to distill data from simulations performed on samples with trimodal and quadmodal texture distributions deformed under plane stress conditions into an adept yield function, and will discuss prediction error. We will conclude with a discussion of the next frontiers.

%%%%%%%%%%%%%%%%%%%%%
\section{Background and Methods}
\label{sec:background}

In this work, we utilize a partially input convex neural network (pICNN) framework to train parameterized yield functions that relate the material microstructure to the resulting macroscopic yield surface. Generally, the training data to the pICNN could be experimentally gathered, simulated, or a combination of both. In this study, we exclusively utilize synthetic data from crystal plasticity finite element (CPFEM) simulations due to their relative ease of generation and low cost as compared to experimental data. In the following subsections, we describe the methods utilized to generate synthetic yield data via CPFEM simulations---including a description of the reduced parameterization of material description---as well as a brief summary of the pICNN framework utilized to train the yield function model.

\subsection{Crystal Plasticity Finite Element Modeling}
\label{subsec:cpfem}

\subsubsection{Model Description}
\label{subsubsec:model}

In this study, we utilize the CPFEM software FEPX~\citep{fepxarxiv,neperfepx} to generate synthetic yield data. FEPX utilizes an elasto-viscoplastic modeling framework embedded in a highly scalable (i.e., MPI-parallelized) non-linear finite element solver. The framework assumes quasi-static, isothermal, ductile material behavior, appropriate for the prediction of macroscopic yield surfaces.

Generally, FEPX considers a domain (i.e., a sample) spatially discretized into grains. A geometry-conforming finite element mesh sub-discretizes each grain into second-order tetrahedral elements, at which the material response is considered. A brief summary of the key components of the kinematics and material model is presented here. For the sake of brevity, we restrict ourselves to a shortened description of the kinematics, material model, and finite element implementation of this well-established method, further details may be found in~\citep{fepxarxiv}.

Regarding the deformation of a material point, we consider the deformation gradient, $\boldsymbol{{\textbf F}}$, to be multiplicatively decomposed into an elastic stretch (the left stretch tensor, ${\textbf V}^{e}$), a rotation (${\textbf R}^{*}$), and a portion describing the plastic deformation (${\textbf F}^{P}$):
\begin{equation}
    \label{eq:def_grad_decomp}
    \boldsymbol{\textbf F} = {\textbf V}^{e} {\textbf R}^{*} {\textbf F}^{P}.
\end{equation}

For the elastic response we relate the Kirchhoff stress, $\boldsymbol\tau$, to the elastic strain, $\boldsymbol{\epsilon}^{e}$, via an anisotropic formulation of Hooke's law:
\begin{equation}
    \label{eq:hooke}
    \boldsymbol{\tau} = {\textbf C({\textbf{r}})} : {\boldsymbol\epsilon}^{e},
\end{equation}
where the fourth-order anisotropic elastic stiffness tensor, ${\textbf C({\textbf{r}})}$, reflects the symmetry of the crystal, and is thus orientation dependent, here written as a function of the Rodrigues parameterization, a vector $\textbf{r}$ ~\citep{frankmrs}. The Kirchhoff stress is proportional to the Cauchy stress via:
\begin{equation}
    \label{eq:stress}
    \boldsymbol{\tau} = \hbox{det}\left({\textbf V}^{e}\right) \boldsymbol\sigma ,
\end{equation}
and that we employ a small elastic strain assumption, such that:
\begin{equation}
    \label{eq:smallelstr}
    {\boldsymbol\epsilon}^{e} = {\textbf V}^{e} - {\textbf I}.
\end{equation}

Regarding plasticity, we first describe the kinetics of slip. We note that equations for plasticity are written in an intermediate configuration/frame, which corresponds to the current/spatial frame with elastic unloading via the inverse of the left Cauchy stretch, ${\textbf V}^{e}$ (see~\citep{Marin1998,Marin1998b}). Here, we employ a rate-dependent power law model, which calculates the shear (slip) rate, $\dot{\gamma}$, on a restricted number of slip systems (chosen based on crystal symmetry):
\begin{equation}
    \label{eq:gamma_dot}
    \begin{split}
        &\dot{\gamma}^{\alpha} = \dot{\gamma}_{0} \left( \frac{\lvert \tau^{\alpha} \rvert}{\tau_c^{\alpha}} \right)^{\frac{1}{m}} \hbox{sgn}(\tau^{\alpha}),
        \\
        \text{where}~&\tau^{\alpha} = \boldsymbol{{\sigma}} : {\textbf P}^{\alpha}.
    \end{split}
\end{equation}
We relate the resolved shear stress, $\tau$, on a slip system, $\alpha$, to the shear rate on that slip system, via the current critical resolved shear stress, $\tau_c$. In this formulation, $\dot{\gamma}_{0}$ is a fixed-state strain rate scaling coefficient and ${m}$ is the fixed-state strain rate sensitivity. The resolved shear stress is calculated as the projection of the stress tensor onto the slip plane and direction, via the symmetric portion of the Schmid tensor, ${\textbf P}$ (itself calculated as the symmetric portion of the tensor formed by the dyadic product between the slip direction vector and the slip plane normal vector~\citep{fepxarxiv}).

We model the evolution of the critical resolved shear stress via an isotropic, saturation-style hardening model:
\begin{equation}
    \label{eq:hardening}
    \dot{\tau_c} =  h_{0} \left( \frac{\tau_s - \tau_c}{\tau_s - \tau_0} \right)\dot{\Gamma}.
\end{equation}
where $h_0$ is the fixed-state hardening rate coefficient, $\tau_s$ is the saturation value of the critical resolved shear stress, $\tau_0$ is the initial critical resolved shear stress, and $\dot{\Gamma}$ is the sum of the shear rates on all slip systems at that material point.

Finally, we update the orientation of the crystal via:
\begin{equation}
    \label{eq:reori}
    \dot{{\textbf r}} = \frac{1}{2} \left( \boldsymbol{\omega} + \left( \boldsymbol{\omega} \cdot {\textbf r} \right){\textbf r} + \boldsymbol{\omega} \times {\textbf r} \right),
\end{equation}
where $\boldsymbol{\omega}$ is based on the plastic spin rate~\citep{fepxarxiv}, a value dependent, ultimately, on the slip system shear rates at that point.

Generally, deformation is applied to the material via the imposition of velocity boundary conditions on the domain of the sample, and a number of time steps are applied at which the solver approximates the solution. Upon completion of the simulation, post-processing may be performed to gain insight into both the evolution of local (i.e., microscale) elasticity and plasticity, but also (of prime importance to this study) the macromechanical deformation response---i.e., the macroscopic stress-strain curve, and ultimately the yield strength of the material. We note that since we utilize CPFEM to calculate the initial (macroscopic) yield surface of the material, the general expectation is that the evolution of the critical resolved shear stresses and the reorientation of the crystals will be relatively low. Indeed, while these values are non-zero throughout the nominally elastic regime, their values are generally negligible at the point of macroscopic yield. We expect the yield surface of the material to be dominated by the microstructure (and particularly the crystallographic texture) of the material and the initial value of the critical resolved shear stress~\citep{Kasemer2017b,Cappola2021}.

\subsubsection{Virtual Sample Generation}
\label{subsubsec:samples}

We perform the CPFEM simulations on high-fidelity representations of microstructures in an effort to understand the relationship between microstructure and resulting macroscopic properties. To achieve this, we must generate both a realistic representation of the geometric morphology of the microstructure and a geometry-conforming finite element mesh, as well as granular orientations that adhere to a target crystallographic texture. We utilize the software package Neper~\citep{Quey2011,neperfepx} for sample generation. 

\begin{figure}[htbp!]
    \centering
    \subfigure[]{%
	\includegraphics[width=0.3\textwidth]{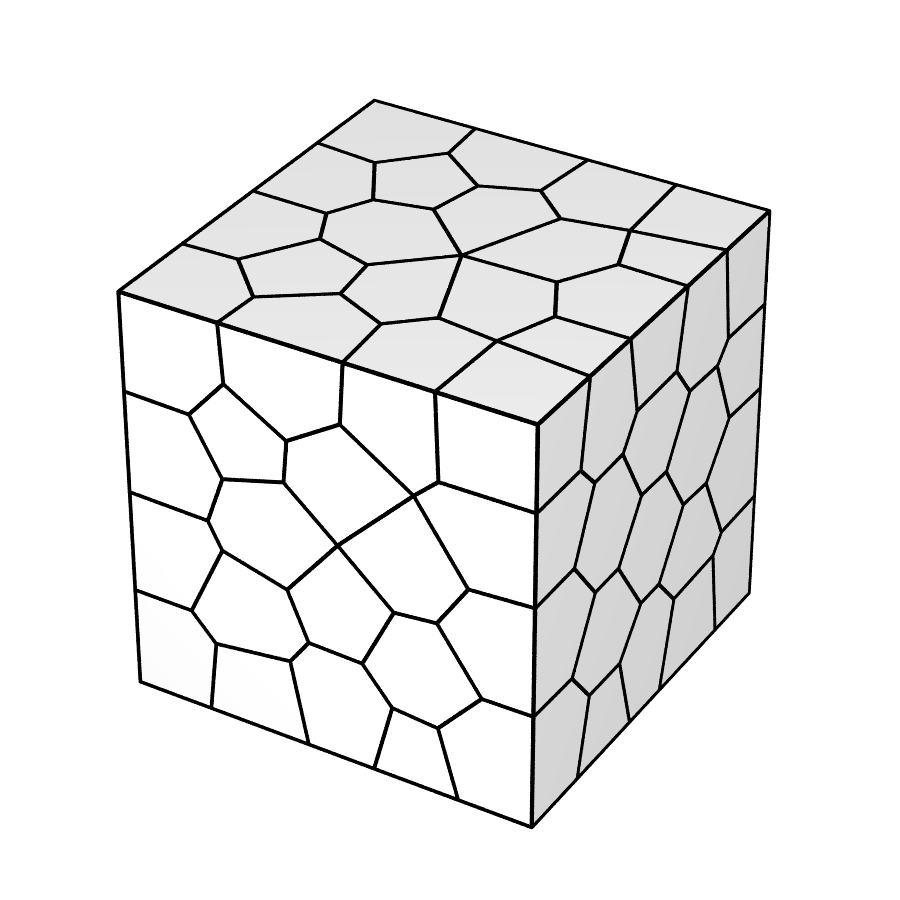}
        \label{subfig:pxgen1}}
    \subfigure[]{%
	\includegraphics[width=0.3\textwidth]{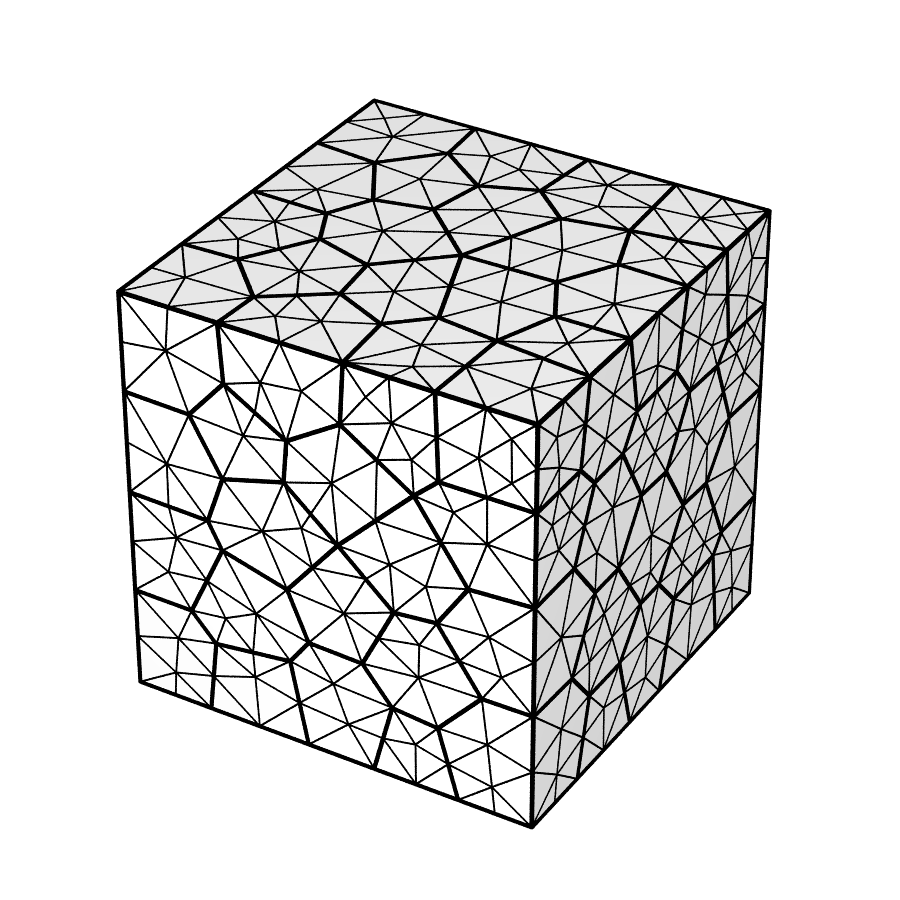}
        \label{subfig:pxgen2}}
    \subfigure[]{%
	\includegraphics[width=0.3\textwidth]{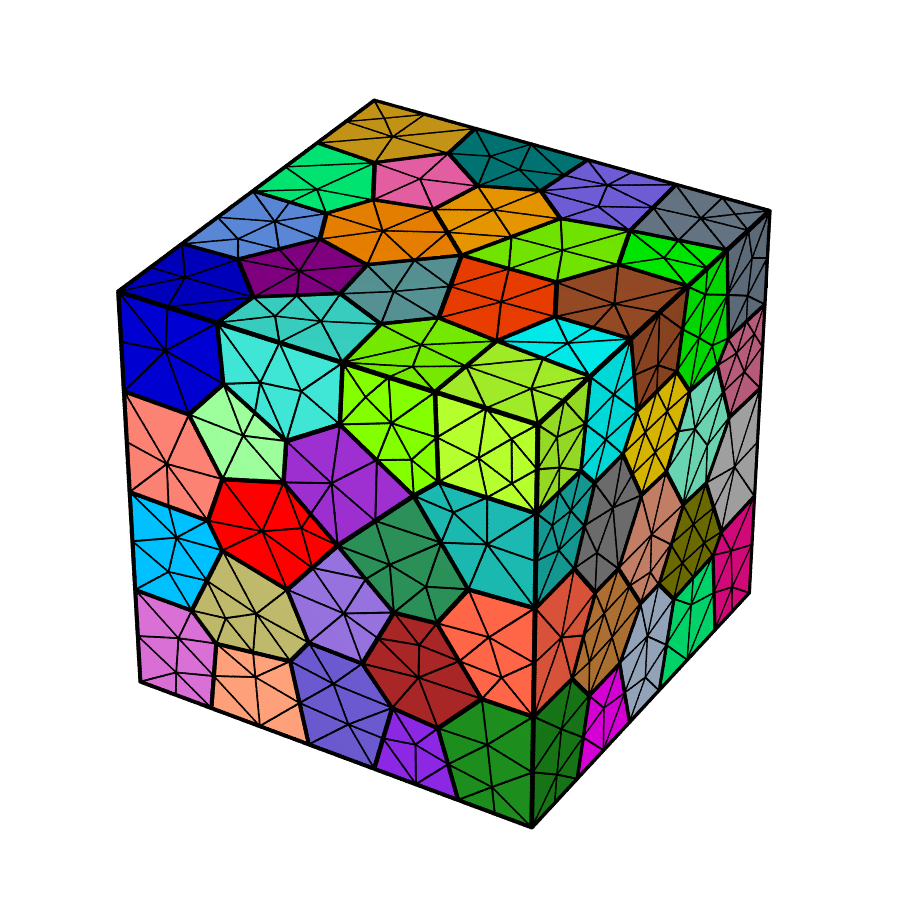}
        \label{subfig:pxgen3}}
    \caption{Example generation of a virtual sample for use in CPFEM simulations, detailing~\subref{subfig:pxgen1} the discretization of a sample domain into regions representing grains,~\subref{subfig:pxgen2} the further discretization of this geometry into a finite element mesh comprised of tetrahedral elements, and~\subref{subfig:pxgen3} the assignment of orientations to the elements within grains (here colored arbitrarily).}
    \label{fig:pxgen}
\end{figure}

Broadly, we present the steps to transition from the representation of the polycrystal domain with explicitly defined grain boundaries (virtual sample) to a finite element mesh with assigned orientations in Figure~\ref{fig:pxgen} for a generic polycrystal. Generally, Neper utilizes Laguerre tessellations~\citep{Kasemer2017b,Quey2018} to generate the geometric morphology of the microstructures (i.e., to discretize the domain into grains), as shown in Figure~\ref{subfig:pxgen1}. We utilize Laguerre tesselations as they allow for the imposition of target distributions of microstructural features---most often the size of grains (e.g., via the distribution of equivalent grain diameters), and the shape of grains (e.g., via the distribution of granular sphericity). We then utilize these tessellations to generate a finite element mesh that conforms to the geometry of the microstructure, as shown in Figure~\ref{subfig:pxgen2}. Finally, we apply orientations to each of the elements within a grain (usually with each grain being, initially, a single-crystal, though this is not generally required), as shown in Figure~\ref{subfig:pxgen3}.

We assign orientations to grains (and by consequence to all elements in each grain) via an approach that considers orientation distribution functions (ODFs). To achieve this, we define ODFs in the fundamental region of Rodrigues space~\citep{frankmrs}. This fundamental region is a closed space whose geometry is determined by crystal symmetry and is an envelope (subset) of orientation space that encompasses the largest set of unique orientations. That is, every orientation that lies outside of the fundamental region is symmetrically equivalent to one lying within the fundamental region. An ODF is a three-dimensional distribution over the fundamental region defining the likelihood of those specific orientations to appear in the sample (essentially acting as a probability density function). Additionally, the integrated mean of the ODF over the fundamental region is 1.

\begin{figure}[htbp!]
    \centering
    \subfigure[]{%
	\includegraphics[width=0.25\textwidth]{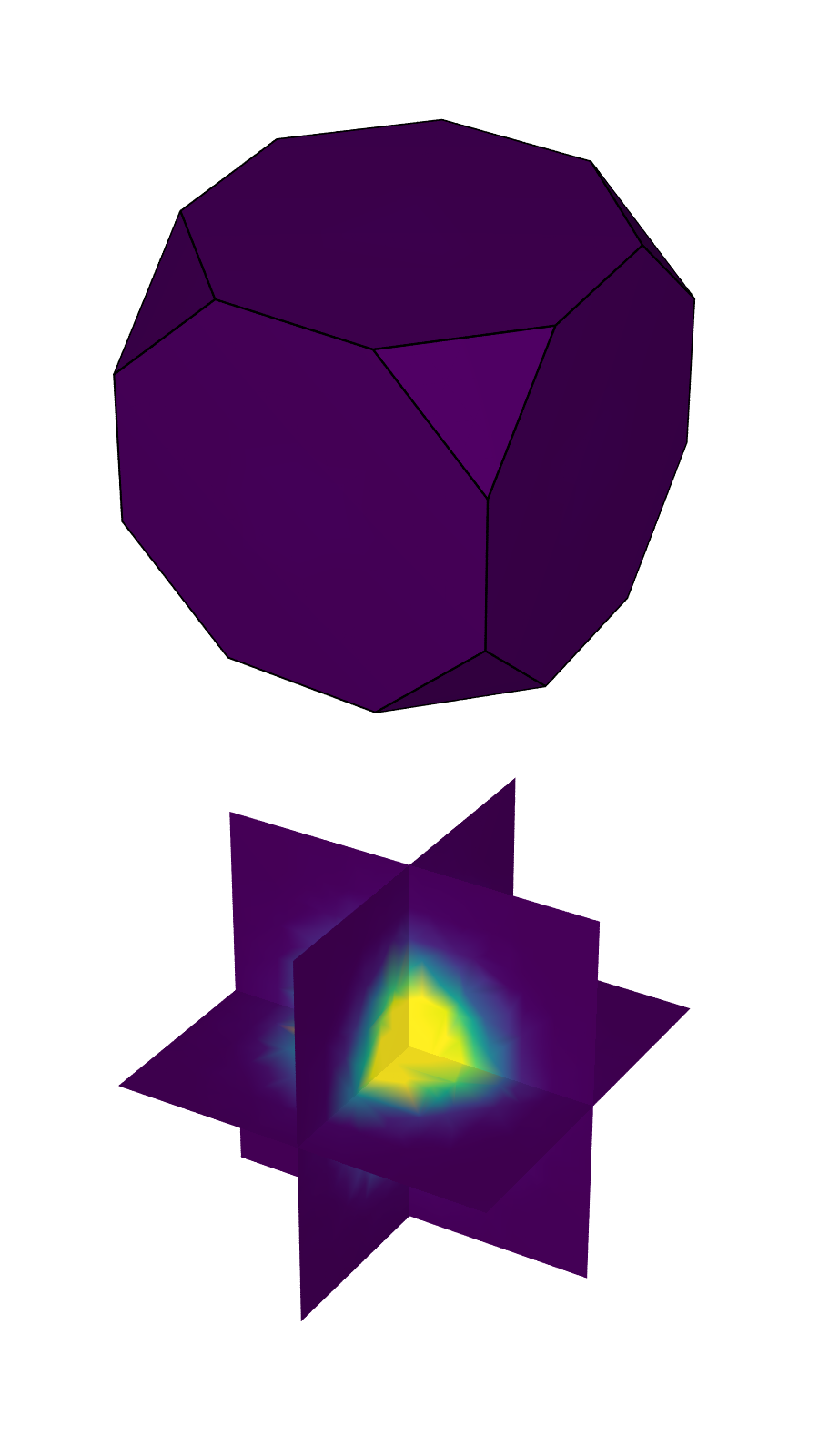}
        \label{subfig:extex1}}
    \subfigure[]{%
	\includegraphics[width=0.25\textwidth]{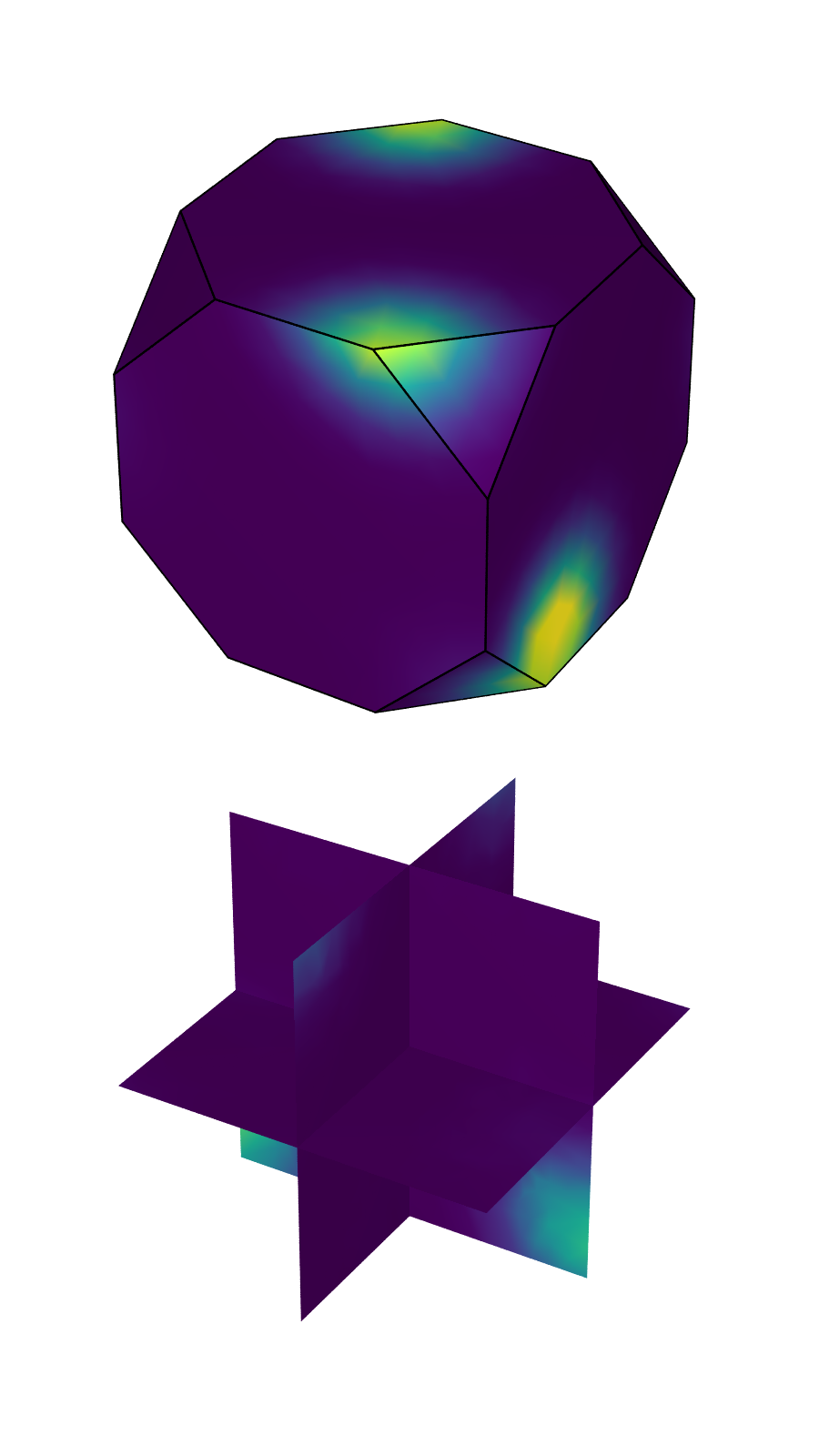}
        \label{subfig:extex2}}
    \subfigure[]{%
	\includegraphics[width=0.40\textwidth]{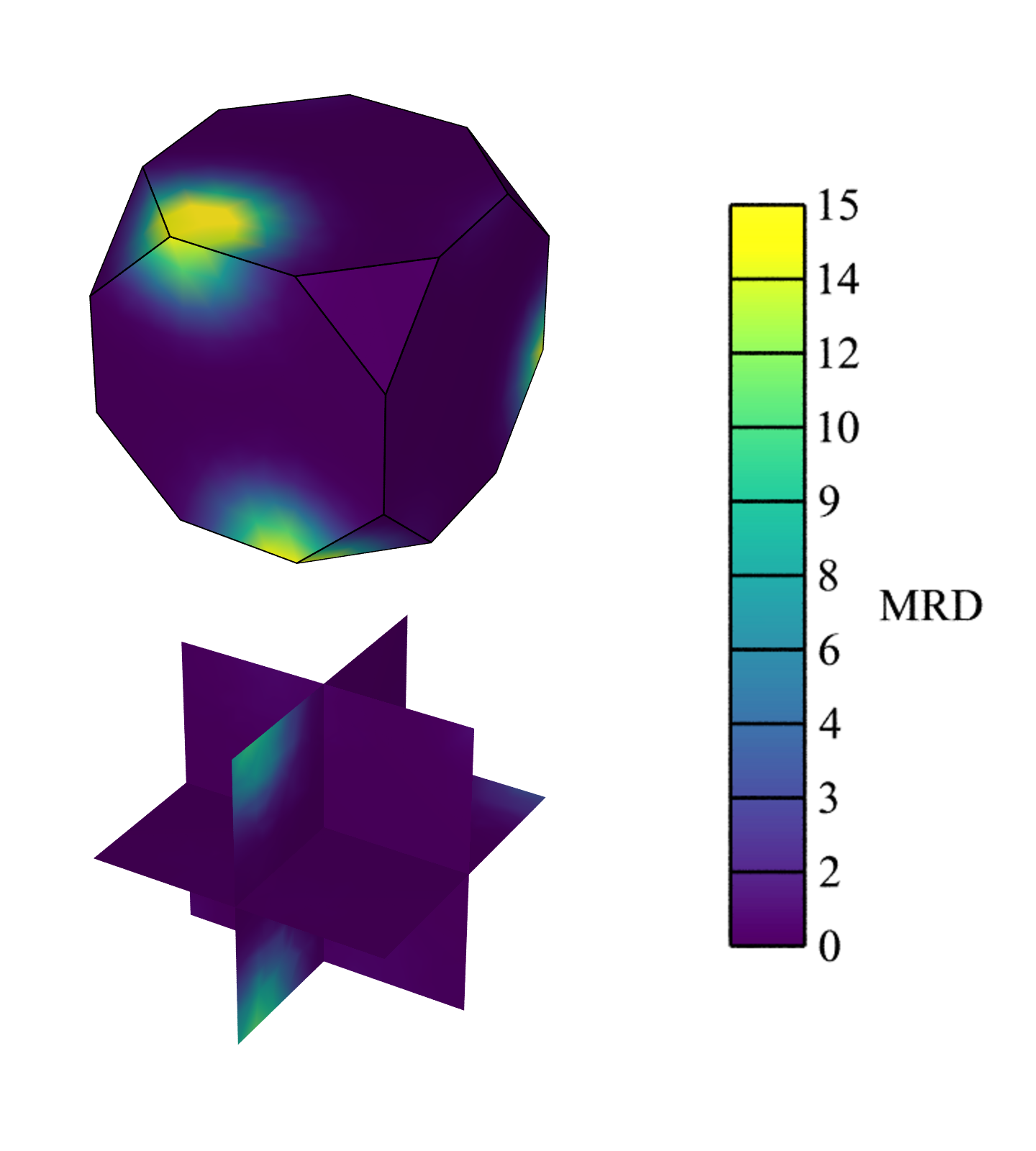}
        \label{subfig:extex3}}
    \caption{Example texture modes for cubic crystals, as plotted on the surface of the cubic symmetry fundamental region (top) and on interior slices (bottom), depicting~\subref{subfig:extex1} a cube texture,~\subref{subfig:extex2} a brass texture, and~\subref{subfig:extex3} a copper texture. Each mode is normalized such that the integrated mean over the fundamental region is 1, and each mode is plotted against the same scale shown in~\subref{subfig:extex3}.}
    \label{fig:extex}
\end{figure}

In this study, we generate synthetic ODFs to mimic realistic textures. For example, crystals exhibiting cubic symmetry processed via known, widely-employed manufacturing routes (i.e., forging, casting, rolling, etc.), tend to exhibit only a handful of common texture modes. These for example include cube texture, brass texture, and copper texture~\citep{backofen,Raabe2005,queyaluminum}. Each of these texture modes is essentially a point-centered distribution in Rodrigues space. Thus, we assume that each texture mode may be represented by a spherical Gaussian distribution centered at these known texture modes, each Gaussian itself is described by its point location in Rodrigues space, as well as a spread~\citep{Raabe2005}~\footnote{We note that some texture modes may not be point Gaussians, but fiber textures. We do not consider such modes in this study, though potentially they may be represented by ``Gaussian cylinders'' in Rodrigues space, (as fibers are linear in Rodrigues space~\citep{queyaluminum}).}. Herein, we quantify the spread of the Gaussian distributions by $\theta$, the arithmetic average of the misorientation angles between the orientations within the distribution and the distribution's average orientation (i.e., the point-center of the distribution). This value is itself related to the 1D standard deviation, $\phi$, of the Gaussian distribution via $\theta = 2\sqrt{2/\pi}\phi$~\citep{glez}. We depict some basic texture modes utilizing a common value of $\theta$ in Figure~\ref{fig:extex}. Overall, we can generally represent textures as the weighted summation of these point textures. We may represent the components of this expansion in vector form:
\begin{equation}
    \label{eq:texvec}
    \mathcal{T} = \{ w_1, \theta_1, w_2, \theta_2, w_3, \theta_3, ..., w_N, \theta_N \},
\end{equation}
where $\mathcal{T}$ is a tuple representing the reduced-order description of texture~\footnote{We recognize the existence of alternative compact methods to represent ODFs, chiefly the expansion of an ODF using discrete spherical harmonics. While spherical harmonics are adept at representing general distributions (and may lead to more realistic representations depending on the ODF), we focus here on physically-realizable textures and thus opt for a more compact reduction by choosing modes which are commonly observed via known processing routes. We do, however, anticipate that the pICNN framework could be trained considering a material description utilizing spherical harmonic weights, albeit with added complexity due to the necessity for a relatively large number of harmonic modes (and thus material descriptors) to accurately reflect texture distributions. In such case one might need to consider additional pre-processing steps for dimensionality reduction.}, $w$ is the weight of the texture mode, and $N$ is the number of texture modes considered. While the texture parameterization may be compactly represented in this manner, to fully reconstruct the texture distribution---and for proper physical interpretation of the texture---the description relies on knowledge of the point location of each texture mode~\citep{queyaluminum} considered (i.e., as depicted in Figure~\ref{fig:extex}).

\subsection{Partially Input Convex Neural Network}
\label{subsec:picnn}

In a previous study~\citep{fuhgcnn}, we have demonstrated the use of partially input convex neural networks (pICNN) as a tool to train surrogate models relating the crystallographic texture to the macroscopic behavior of the material in a thermodynamically consistent manner by enforcing convexity of the yield function with regards to the six-dimensional Cauchy stress space, i.e. Drucker's hypothesis~\citep{drucker1952more}. In this previous study, however, we focused on an irreducibly simplified problem of a single-component texture. While the pICNN framework was proven adept at predicting macroscopic yield surfaces for this relatively simple problem, we did not explore more complex material states (i.e., complex crystallographic textures), nor construct the pICNN architecture to properly handle these states. Here, we will briefly summarize the basic features of the pICNN framework, as well as the extensions made to allow for the training of more complex material states. Overall, we refer the reader to~\citep{fuhgcnn} for a full description of the pICNN framework.

\begin{figure}
    \centering
    \includegraphics[width=3.15in]{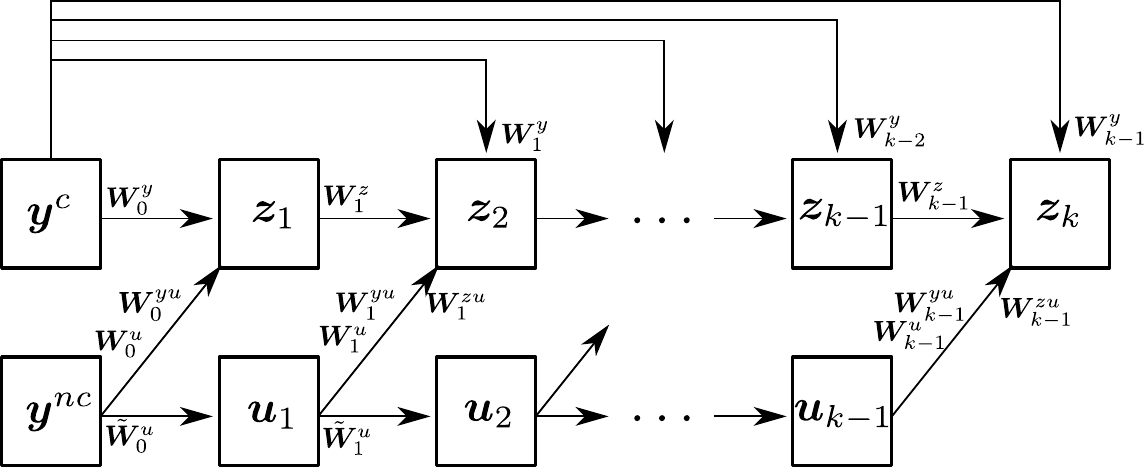}
    \caption{Schematic depicting the basic architecture of the partial input convex neural network (reproduced from~\citep{fuhgcnn}).}
    \label{fig:picnn}
\end{figure}

We utilize a typical neural network that contains one input layer, multiple hidden layers, and one output layer. We assume that each hidden layer consists of a set of neurons. To ensure convexity with respect to all input, Amos et al.~\citep{amos2017} formulated a modified update scheme based on typical neural network formulations. Of note, this scheme works, in part, by choice of a convex and non-decreasing activation function, as well as enforcement of non-negative weights in the update scheme. In our previous study, also following Amos et al.~\citep{amos2017}, we aimed to enforce convexity only for a subset of the input data. Namely, we did not require the neural network to be convex with regard to the texture descriptors, but only to the stresses. We present a schematic of the pICNN architecture in Figure~\ref{fig:picnn}.

To train the pICNN, we require a description of the material and the resulting yield behavior, such that the algorithm can generate a yield function that relates the microstructure of the material to its resulting macroscopic properties. Focusing on the description of the material, we define a material description vector, $\mathcal{M}$, which may contain variables describing the material state: possibly including descriptors of the material behavior (i.e., material parameters), the geometric morphology (e.g., via, perhaps, the statistics of the grain size and shape distributions, among other metrics), and the crystallographic texture. While a goal for future studies, here, we aim to establish a relationship between complex crystallographic textures alone and the macroscopic yield of the material. Consequently, we choose to limit our material description vector to describe only the crystallographic texture, or $\mathcal{M} = \mathcal{T}$ (Equation~\eqref{eq:texvec}).

In our previous study, the material description vector was univariate. As we considered only one texture mode, we were able to wholly describe the crystallographic texture via this mode's spread, or $\mathcal{M} = \theta$, as this mode's weight was necessarily 1, and all other modes 0. Extending the ability to consider multimodal texture distributions requires a more complex material description vector (shown in Equation~\eqref{eq:texvec}). Consequently, we must modify the architecture of the pICNN that we had previously developed to scale with the complexity of the material description vector. To achieve this, we consider the size of the array of training data to determine the number of hidden layers in the algorithm. 

Overall, the size of the training data array is dependent on the complexity of the material description vector as well as the size of the grid in stress-space over which the yield surface is defined (see:~\citep{fuhgcnn} for a description of our strategy). For this study, we consider a regular grid in stress-space with $N_{grid} = 103,041$ points. In general, according to material description vector considered in this study, we have $2N$ material descriptors, as well as the grid points in stress space (two-dimensional, when considering plane-stress problems). Thus, our total training data array is of shape $(2N+2) \times N_{grid}$ (again, previously the shape of the training data array was limited to $3 \times N_{grid}$ as the material description was univariate). The output vector is of length $N_{grid}$, which contains the equivalent stresses at the grid point, which is ultimately used to determine whether a grid point is elastic or plastic (i.e., to determine the level set that is the yield surface). We performed a brief study to ensure the model does not overfit, and ultimately chose three hidden layers sized $2N \times 10$, $2N \times 10$, and $2N$. We choose the activation function (Adam \citep{kingma2014adam}), weight initialization (Glorot's uniform distribution \citep{glorot2010understanding}), loss function (mean squared error), and learning rate ($10^{-4}$) to match those from our previous study. We perform no rigorous hyperparameter study to optimize the training efficiency (it is not within the scope of this study, and we reserve this topic for future work).

%%%%%%%%%%%%%%%%%%%%%
\section{Crystal Plasticity Simulations}
\label{sec:Simulations}

Under the above assertion that our prime goal in this study is to explore the relationship between complex textures and the macroscopic response, we will hold many of the features of the samples fixed while varying texture. Again, we note that this is not motivated by any perceived limitation of the pICNN framework, and we expect the framework to be adept at handling further additions to the material description vector with an increase in training data requirements. We reserve these additions for future studies, and thus consideration of geometric morphology or crystal behavior in the pICNN is not within the scope of this study. Overall, the description below thus focuses on the generation of a single virtual specimen over which the crystallographic texture is varied.

\subsection{Material Parameters}
\label{subsec:simmaterial}

We opt to utilize Okegawa mold copper (OMC) as an appropriate test material for this study. OMC copper exhibits cubic symmetry, as well as appreciable elastic anisotropy (Zener ratio of 3.2) which will lead to pronounced differences in macroscopic yield surfaces as a function of changes to crystallographic texture. We utilize the elastic/plastic parameters from~\citep{obstalecki2014,wong2015} in this study, as summarized in Tables~\ref{tab:elastic} and~\ref{tab:plastic}. We again note that---since we are primarily interested in the point of the initial yield of the sample---the choice of the initial slip system strength, $\tau_0$, is important, but the choice of other plasticity parameters is largely arbitrary (within reason), as they primarily affect the evolution of plasticity, which is not expected to be large at the point of macroscopic yield.
\begin{table}[htbp!]
    \centering
    \begin{tabular}{c c c}
    \hline
    {\bf $C_{11}$ (\SI{}{\giga\pascal})} & {\bf $C_{12}$ (\SI{}{\giga\pascal})} & {\bf $C_{44}$ (\SI{}{\giga\pascal})} \\
    \hline
    164 & 122 & 75 \\
    \hline
    \end{tabular}
    \caption{Single crystal elastic constants as found in~\citep{wong2015}.}
    \label{tab:elastic}
\end{table}

\begin{table}[htbp!]
    \centering
    \begin{tabular}{c c c c c}
    \hline
    {\bf $\dot{\gamma}_0$ (-) } & {\bf $m$ (-)} & {\bf $h_0$ (\SI{}{\mega\pascal})} & {\bf $\tau_{0}$ (\SI{}{\mega\pascal})} & {\bf $\tau_s$ (\SI{}{\mega\pascal})} \\
    \hline
    1 & 0.01 & 800 & 85 & 285 \\
    \hline
    \end{tabular}
    \caption{Plastic modeling parameters as found in~\citep{wong2015}.}
    \label{tab:plastic}
\end{table}

\subsection{Sample Generation}
\label{subsec:sample}

Referring to our general description of sample generation in Section~\ref{subsubsec:samples}, we begin by noting that we wish to generate a sample with relatively equiaxed grains---i.e., grains of near-equal size and shape. In this way, we intend to mute the effect that the geometric morphology of the grains has on the deformation response of the material (i.e., as might be expected from highly elongated grains, or other such appreciably anisotropic structures). As such, we choose Dirac distributions of value 1 for both the distribution of grain size and the distribution of sphericity (i.e., a ``centroidal'' tessellation), as the lack of distribution spread facilitates grains with near equal size and shape metrics. We choose to generate a sample with 100 grains, as this proves sufficient for producing consistent macroscopic yield predictions (i.e., multiple samples generated containing the same nominal texture but different individual orientations predict yield strengths within 5\% of the average~\citep{fuhgcnn}). We choose to mesh the sample with approximately 10,000 elements (i.e., 100 elements per grain), as this likewise proves sufficient in providing consistent macroscopic predictions. The geometric morphology utilized in this study is shown (previously) in Figure~\ref{fig:pxgen}.

\subsection{Crystallographic Texture}
\label{subsec:texture}

As we wish to explore the effect that complex textures have on the macroscopic yield surface, we generate several samples with varying mixtures of multimodal textures. In this study, we choose to limit the texture-weight space that we explore to mixtures of three different texture modes: cube, brass, and copper (presented in Figure~\ref{fig:extex}). Consequently, our material description vector becomes:
\begin{equation}
    \mathcal{M} = \{ w_c, \theta_c, w_b, \theta_b, w_{Cu}, \theta_{Cu} \},
\end{equation}
where subscript $c$ refers to the cube texture mode, $b$ to the brass mode, and $Cu$ to the copper mode. 

To generate the texture mixtures for use in either training or testing, we assume that the sum of the weights must be 1 (or 100\%)---i.e., that we contain no other texture modes but cube, brass, and copper. Assuming that all weights are positive, and for a fixed value of $\theta$, we can thus represent any available texture on a triangular plane (i.e., described by the equation $w_c + w_b + w_{Cu} = 1$) in 3D texture-weight space, whose bases are the weights of the cube, brass, and copper modes. This triangular plane may, alternatively, be represented as a ternary diagram, which we will utilize exclusively.

\begin{figure}
    \centering
    \subfigure[]{%
	\includegraphics[width=2.1in]{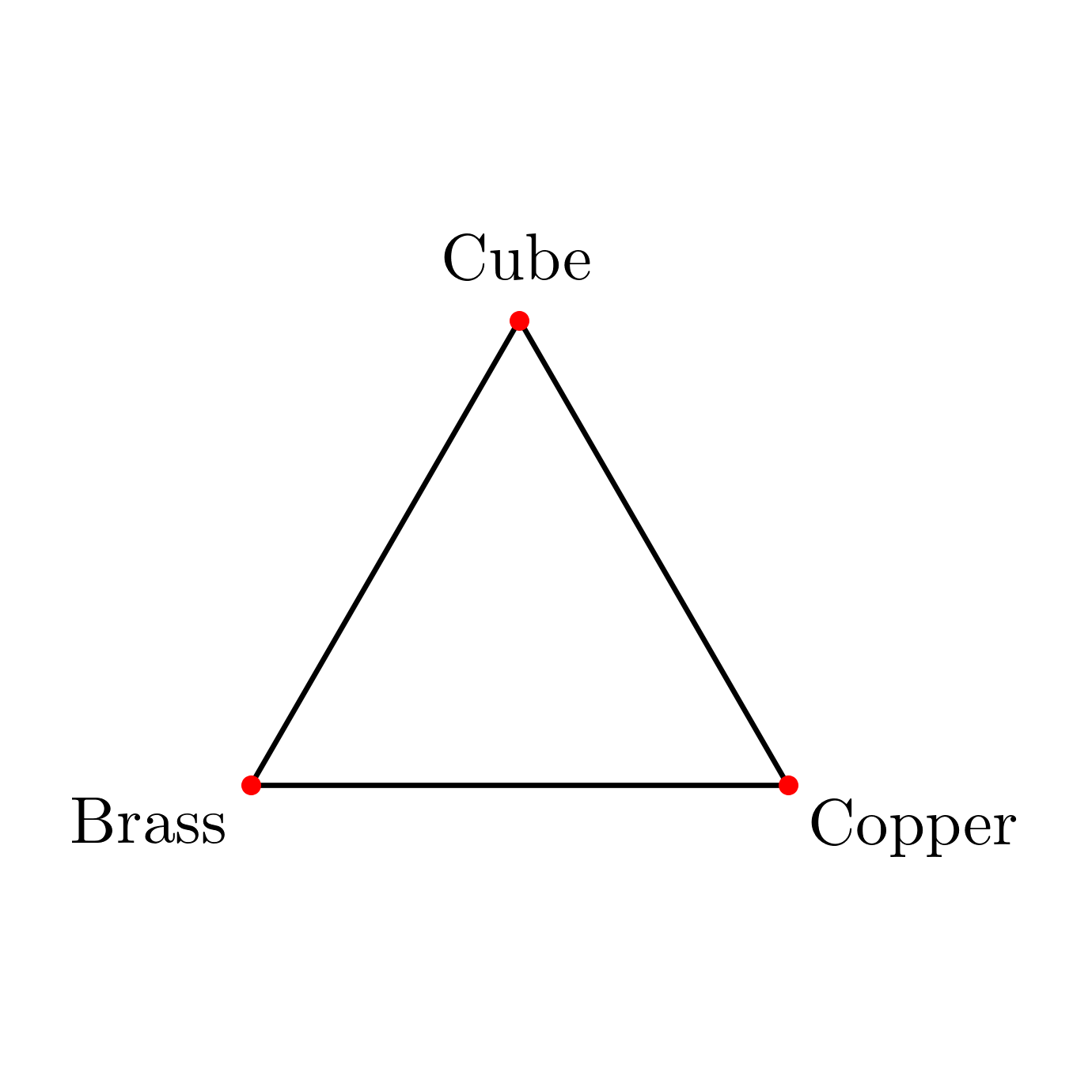}
        \label{subfig:disclevel0}}
    \subfigure[]{%
	\includegraphics[width=2.1in]{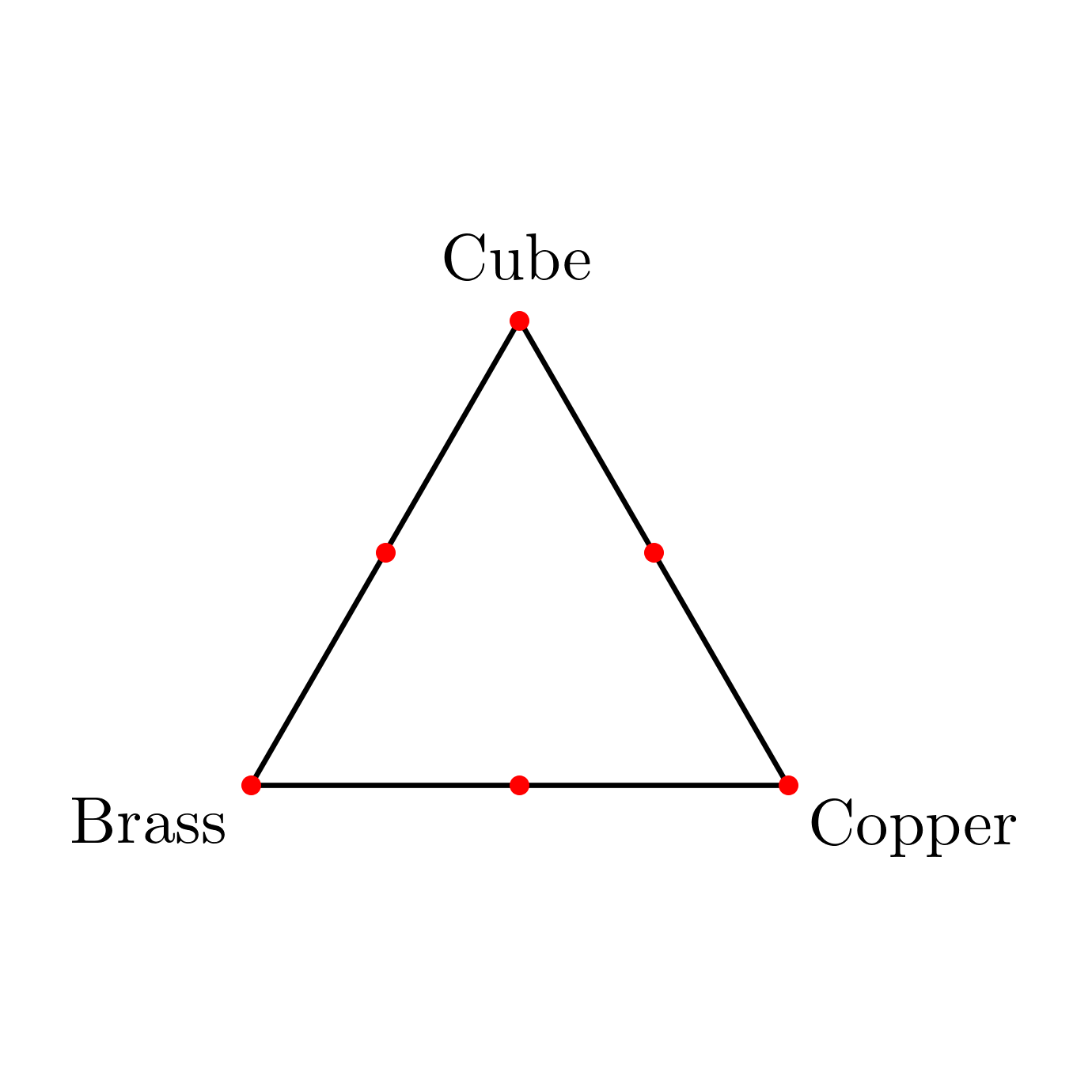}
        \label{subfig:disclevel1}}
    \subfigure[]{%
	\includegraphics[width=2.1in]{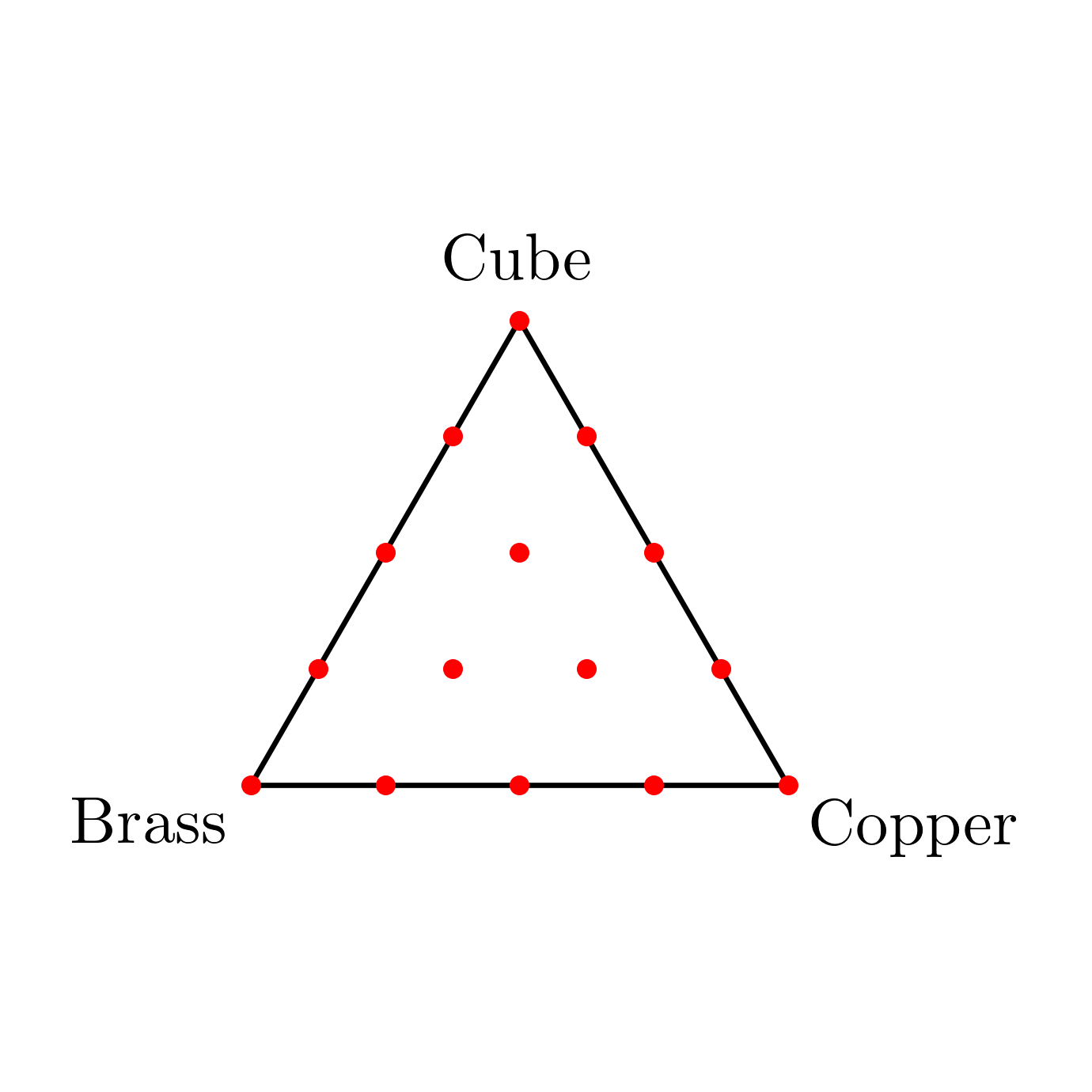}
        \label{subfig:disclevel2}}    
    \caption{Example discretization of a plane in trimodal texture-weight space, considering the cube, brass, and copper point-texture modes, where the sum of the modal weights is 1, specifically \subref{subfig:disclevel0} level 0 discretization, \subref{subfig:disclevel1} level 1 discretization, and \subref{subfig:disclevel2} level 2 discretization.}
    \label{fig:disclevels}
\end{figure}

\begin{figure}
    \centering
\includegraphics[width=3.15in]{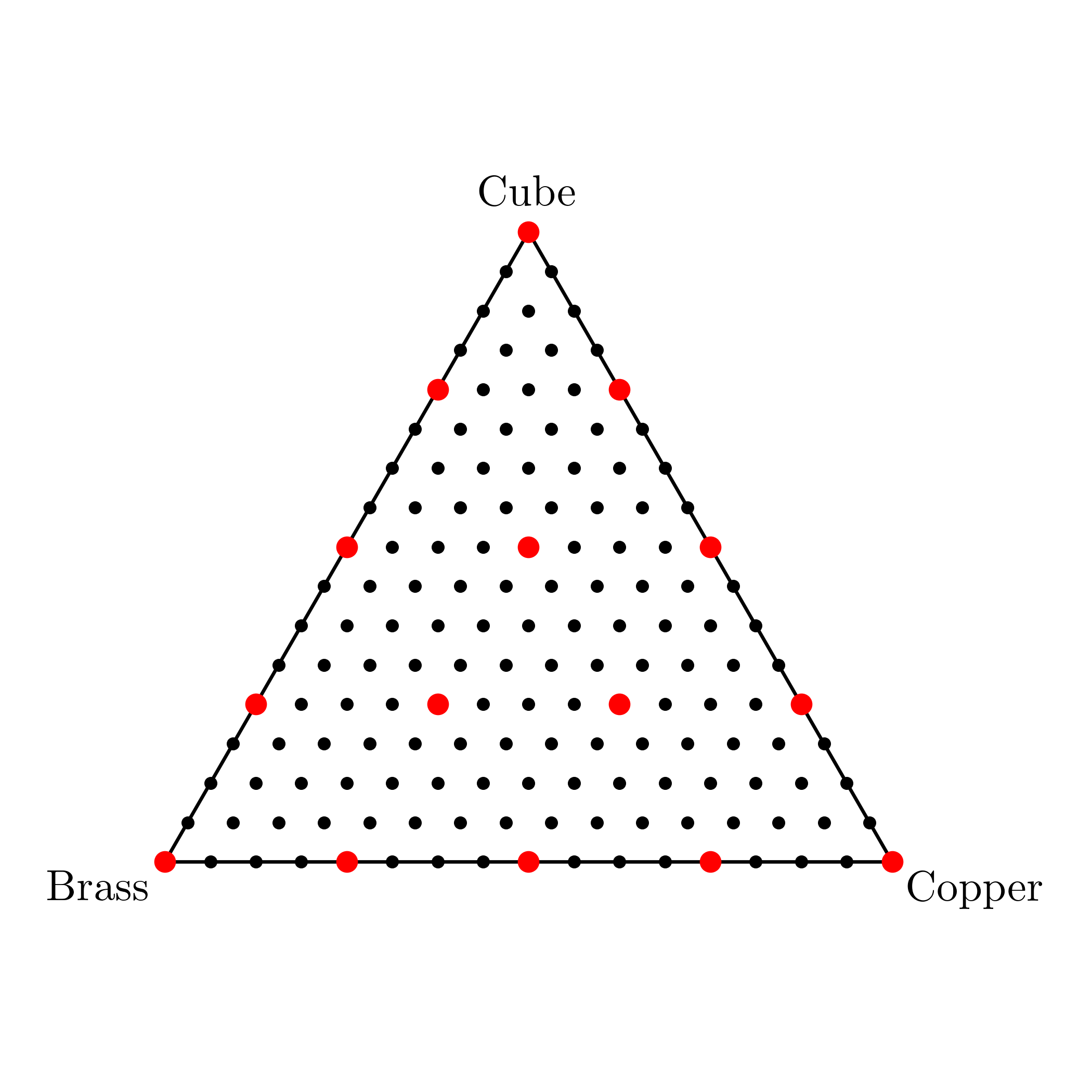}
    \caption{A ternary plot, where all points plotted represent textures/samples at which CPFEM simulations are performed for either training or test cases. The red points presented represent the textures utilized for generating training data, while black points represent textures to be utilized for testing.}
    \label{fig:textri}
\end{figure}

To generate an even density of points (which each represents a sample/texture mixture) across the ternary diagram for training and testing purposes, we utilize a recursive discretization scheme, as represented in Figure~\ref{fig:disclevels}. Here, we consider different levels of discretization, where the 0th level indicates points only at the vertices of the ternary diagram (only unimodal textures, Figure~\ref{subfig:disclevel0}), the 1st level of discretization contains, additionally, the midpoints of the triangle formed by the vertices of the diagram (adding bimodal textures, Figure~\ref{subfig:disclevel1}), and the 2nd level of discretization contains, additionally, the midpoints of the triangles formed by these points (considering trimodal textures, Figure~\ref{subfig:disclevel2}). This can be repeated for any level of desired discretization. The samples that we will utilize for training the pICNN model are shown as red points in Figure~\ref{fig:textri} (level 2 discretization, 15 samples total), while the test samples are shown as black points in the same figure (level 4 discretization, 153 samples total). We note that the training samples and test samples intersect.

\subsection{Boundary Conditions}
\label{subsec:simbcs}

For each crystallographic texture described in the previous section, we perform a number of simulations to elucidate the yield surface for that given texture/sample. Generally, we apply normal loads to the surfaces of the domain of the polycrystal. In this study, we choose to focus on biaxial loading and, consequently, plane-stress yield surfaces in an effort to limit computational cost. For each simulation performed, we apply loading where the biaxial ratio (e.g., the ratio of $\sigma_z$ to $\sigma_y$, the applied stresses in the $z$ and $y$ directions) is held fixed through deformation---i.e., the deformation path is a constant radial vector in biaxial stress-space, which we will herein refer to as ``load vectors''. For a given simulation performed along a load vector, we apply deformation with a sufficient number of time steps in an effort to accurately capture the point of yield in the sample, as determined via a 0.1\% offset of the equivalent stress vs. equivalent strain curve.

\subsection{Strategy for Yield Surface Data Generation}
\label{subsec:adaptive}

To generate the biaxial yield surface for a given texture, we perform multiple simulations on the same sample. As the CPFEM simulations are the most computationally costly portion of this workflow, we seek to limit the number of simulations necessary to accurately reflect the shape of the yield surface. We assume no tension-compression asymmetry~\citep{fuhgcnn} and as such use the results from simulations in the first and second quadrants to fill the third and fourth quadrants. By simulating deformation paths along vectors only in the first and second quadrants, we reduce the number of load vectors necessary to elucidate the yield surface by (nominally) half. 

\begin{figure}
    \centering
    \subfigure[]{%
	\includegraphics[width=3.15in]{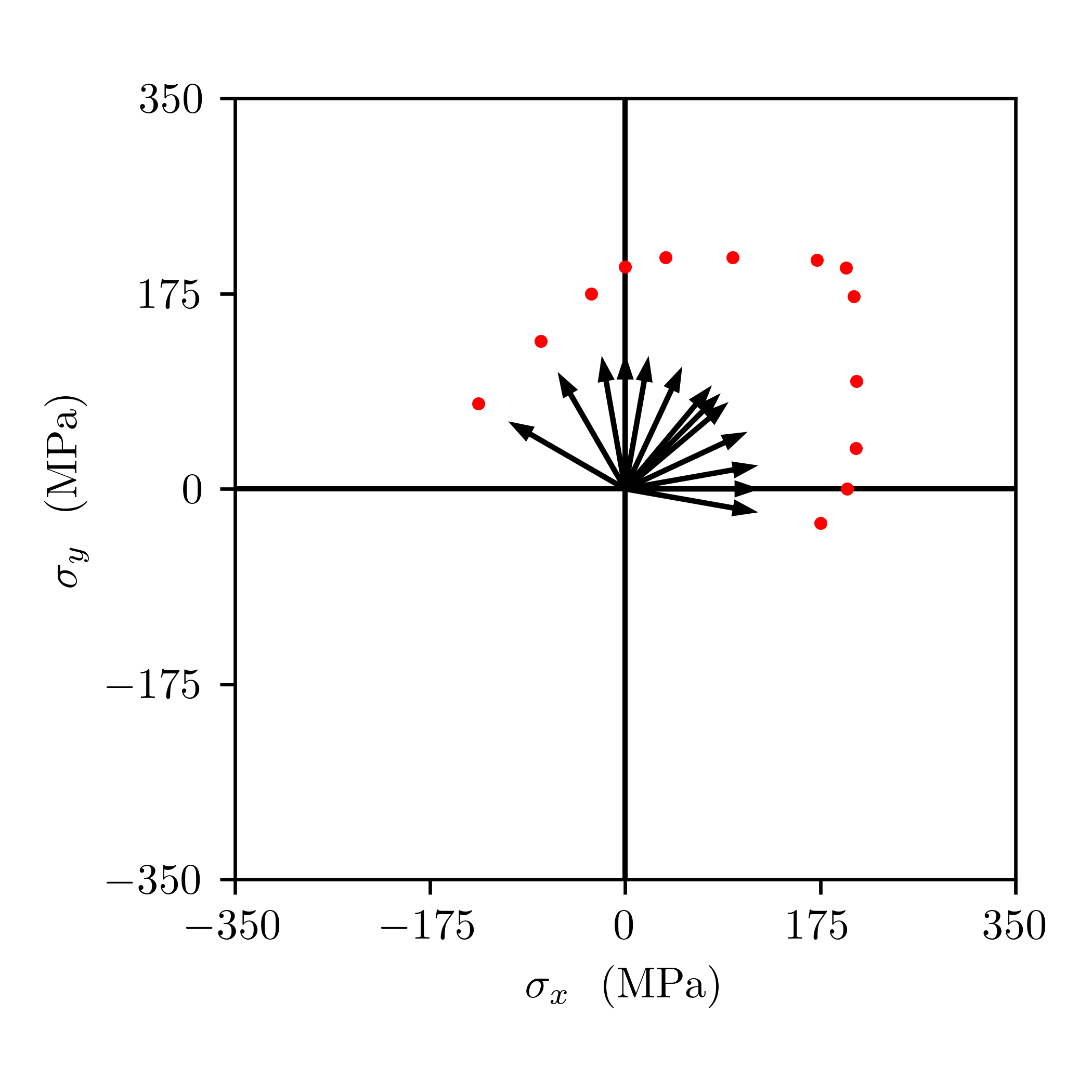}
        \label{subfig:exloadvecs1}}
    \subfigure[]{%
	\includegraphics[width=3.15in]{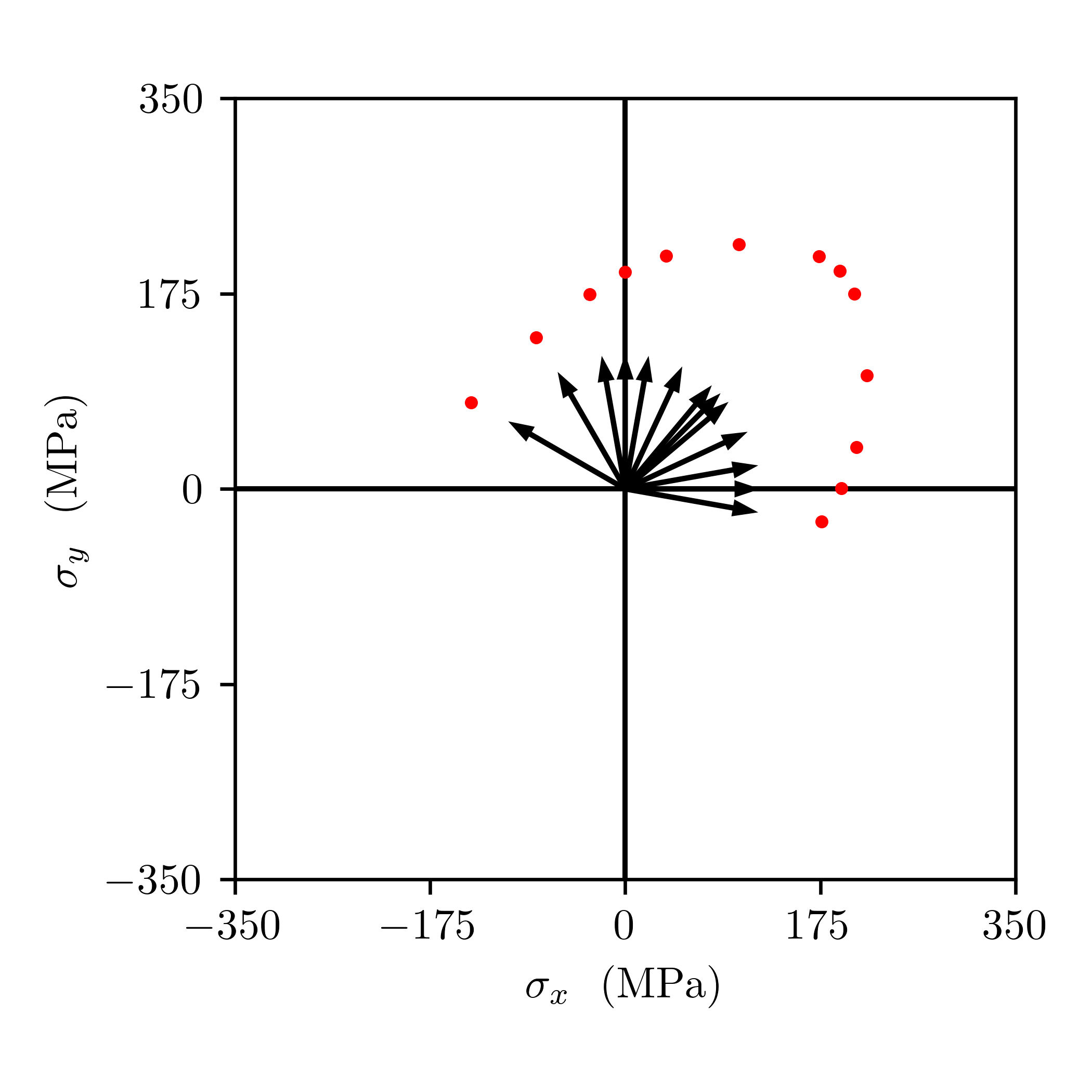}
        \label{subfig:exloadvecs2}}
    \caption{Example load vectors (black lines) and resulting yield points along these vectors (red points) for a set of example simulations with~\subref{subfig:exloadvecs1} a sample with strong texture (i.e., highly faceted), and~\subref{subfig:exloadvecs2} a sample with weak texture (i.e., weakly faceted). Note the placement of the vectors near the vertex of the yield surface in the first quadrant, whose directions are determined from the results of the uniaxial simulations.}
    \label{fig:exloadvecs}
\end{figure}

To further limit the number of load vectors, we strategically choose their placement in the first and second quadrants to reflect expectations regarding the shape of the yield surface. In essence, we wish to place load vectors that elucidate the positions of the vertices of the yield surface, as well as points near the vertices to give a sense of how faceted the yield surface is. To achieve this, we perform uniaxial tests in both the $x$ and the $y$ direction. We then utilize the results from the uniaxial tests to estimate the location of the vertex of the yield surface in the first quadrant---which generally tends to be along a load vector with the ratio of the yield stresses from the uniaxial simulations. We adaptively place three narrowly spaced load vectors near the location of this expected vertex (i.e., \SI{5}{\degree} spacing azimuthally in stress-space), along with two enveloping load vectors near the uniaxial cases to elucidate these vertices (\SI{10}{\degree} spacing azimuthally in stress-space, note that one vector falls into the fourth quadrant), as well as two further load vectors in the first quadrant (\SI{15}{\degree} from the uniaxial groupings) and two other widely-spaced load vectors in the second quadrant (each \SI{30}{\degree} from the axes) to aid in capturing curvature. Thus, we utilize 13 load vectors to elucidate the shape of the yield surface, an example of which is shown for two samples in Figure~\ref{fig:exloadvecs} (previously, we utilized results from 72 CPFEM simulations performed along equally-spaced load vectors to elucidate the shape of the yield surface). For the 153 different textures/samples that we consider as described in Section~\ref{subsec:texture}, we thus perform a total of 1,989 simulations to elucidate the yield surfaces for all samples.

\subsection{Simulation Post-processing}
\label{subsec:dataproc}

With the important aspects of the yield surface (i.e., vertices and degree of curvature/facetedness) determined via results from the strategically-placed CPFEM simulations, we turn to interpolation to fill the spaces in between the load vectors at which simulations were performed~\citep{fuhgcnn}. The interpolated data will then uniformly sampled and directly utilized for the training of the pICNN framework. Interpolation is necessary such that the pICNN has a high-dimensional representation of the yield surface (i.e., no large gaps between load vectors, such as what is present in the raw CPFEM data presented in Figure~\ref{fig:exloadvecs}). We choose to interpolate, as it is computationally significantly cheaper than performing a wide array of CPFEM simulations. 

Because the simulation data is not exact due to convergence tolerances, the points along the yield surface do not necessarily create a convex hull, and consequently, in the case of interpolation the yield surface may not be a ``closed-loop''. In other words, we generally have a surface/curve that contains two numerically close endpoints, that are not coincident (euclidean distance on the order of \SI{1e-5}{\mega\pascal}), which presents minor challenges when interpolating between the points. We construct multiple B-spline curve fits~\citep{rogers1989constrained,piegl2012nurbs} among all 26 data points (i.e., the 13 simulations and their negatives) with tangency conditions at the ends to maintain curve continuity, a consequence of the above-explained dis-coincidence. We consider piece-wise first-order, second-order, and third-order B-spline fits curves. We combine these curves into a single spline considering a weighted average of the functions where the weights are selected to minimize the error across a variety of yield curves (i.e., both highly-faceted and not). The splines thus functionally define the yield curve based on the limited CPFEM data points. From this functional definition, we then calculate any number of points desired that themselves represent a discretized yield curve, though with whatever desired density of points and spacing in stress space. These points are then as training data for the pICNN, as its purpose is to signal the yield behavior.

\begin{figure}
    \centering
    \subfigure[]{%
	\includegraphics[width=3.15in]{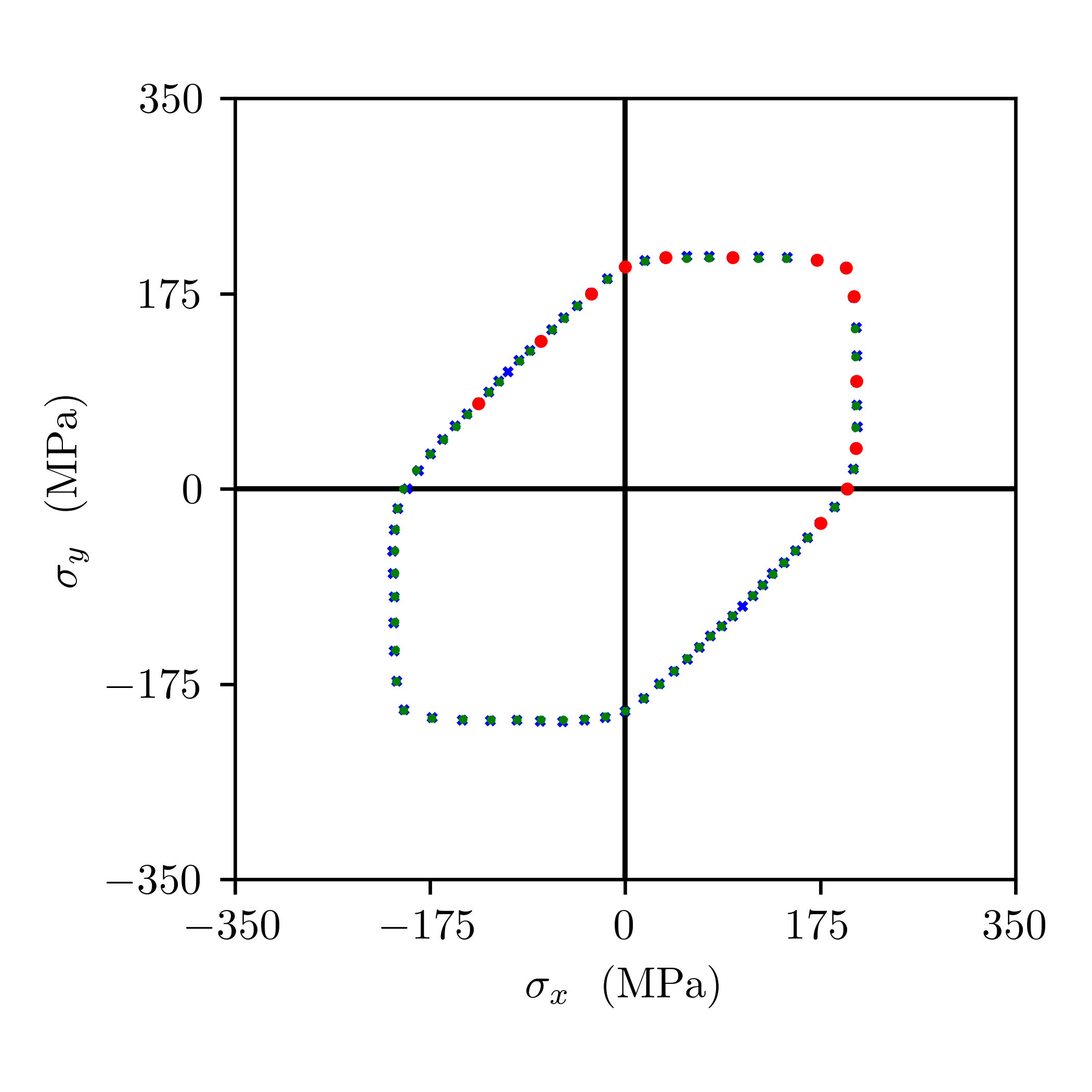}
        \label{subfig:exinterp1}}
    \subfigure[]{%
	\includegraphics[width=3.15in]{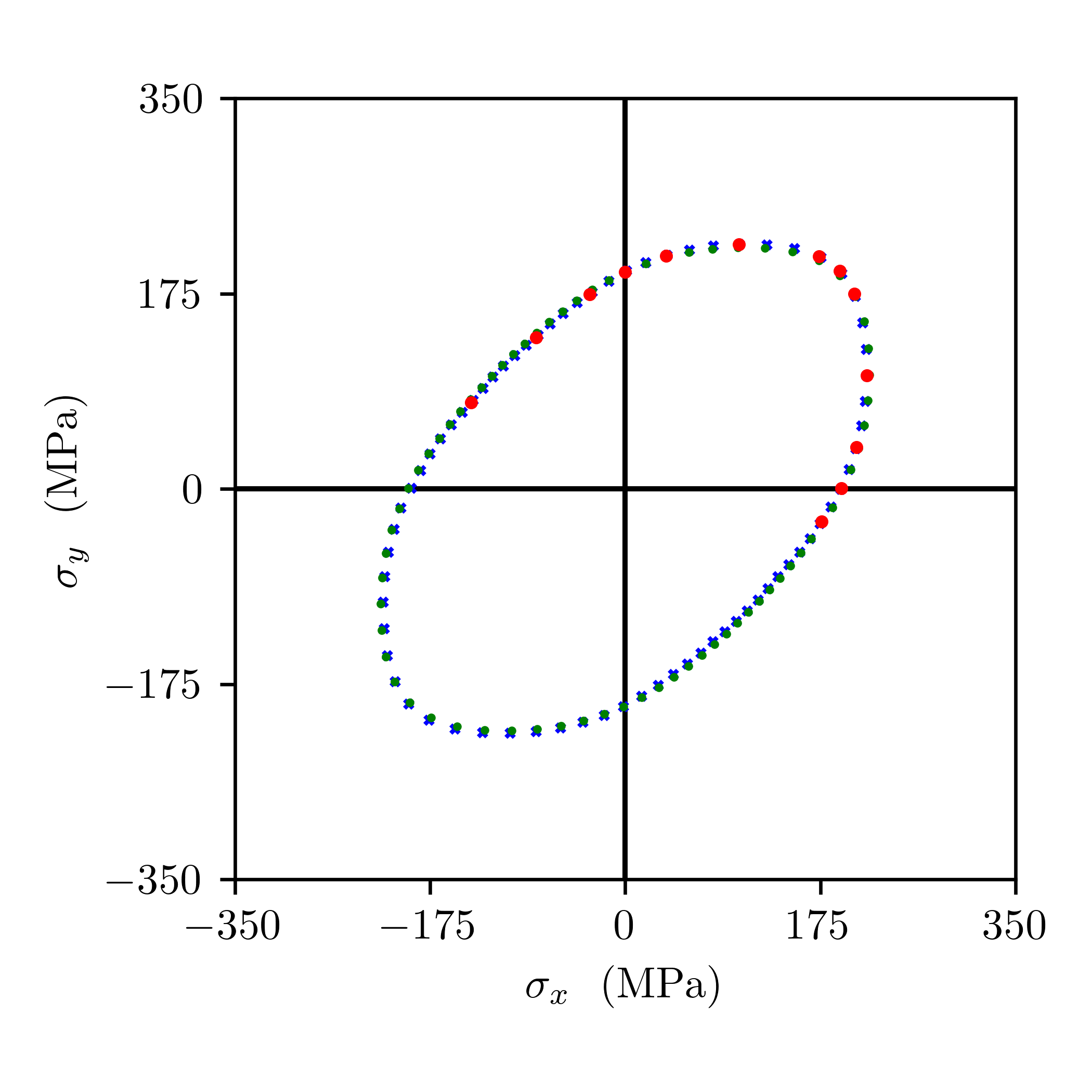}
        \label{subfig:exinterp2}}
    \caption{Example results from interpolation (blue points) between the reduced-set of CPFEM results (red points), as well as the actual results from the CPFEM simulations along the same load vectors as the interpolated points (green points) for a set of example simulations with~\subref{subfig:exinterp1} a sample with strong texture (i.e., highly faceted), and~\subref{subfig:exinterp2} a sample with weak texture (i.e., weakly faceted).}
    \label{fig:exinterp}
\end{figure}

We demonstrate the interpolation method by employing it on the yield points for the two demonstrative samples in Figure~\ref{fig:exloadvecs}, and present the results in Figure~\ref{fig:exinterp}. Here, we interpolate along 72 equally-spaced load vectors (i.e., \SI{5}{\degree} spacing) in biaxial stress space, utilizing the 13 CPFEM points to guide the interpolation. We further plot the CPFEM results along these same 72 load vectors to demonstrate the acceptable results of the interpolation. We find that the mean percentage error between the interpolated points and the actual CPFEM points is 0.38\% for the case of the strong texture ($\theta=$~\SI{5}{\degree}), and 0.97\% for the case of the weak texture ($\theta=$~\SI{25}{\degree}). This indicates that our interpolation scheme is broadly acceptable in elucidating the proper shape of the yield surface given the 13 yield points we calculate from CPFEM, including both on highly faceted and curved surfaces. In summary, by employing interpolation, we are able to confidently and accurately provide the same density and spacing of yield points in biaxial stress-space as in our previous study (i.e., 72 vectors) with only 13 strategically-placed CPFEM vectors total---i.e., a more than five-fold reduction in the number of CPFEM simulations (the most costly portion of the workflow) necessary to elucidate the yield surface. The strategic choice of the 13 CPFEM load vectors facilitates this method, as fidelity around the vertices would be lost if interpolation were performed on randomly placed vectors.

%%%%%%%%%%%%%%%%%%%%%

\section{Results}
\label{sec:Results}

\begin{figure}
    \centering
    \subfigure[]{%
        \includegraphics[width=3.15in]{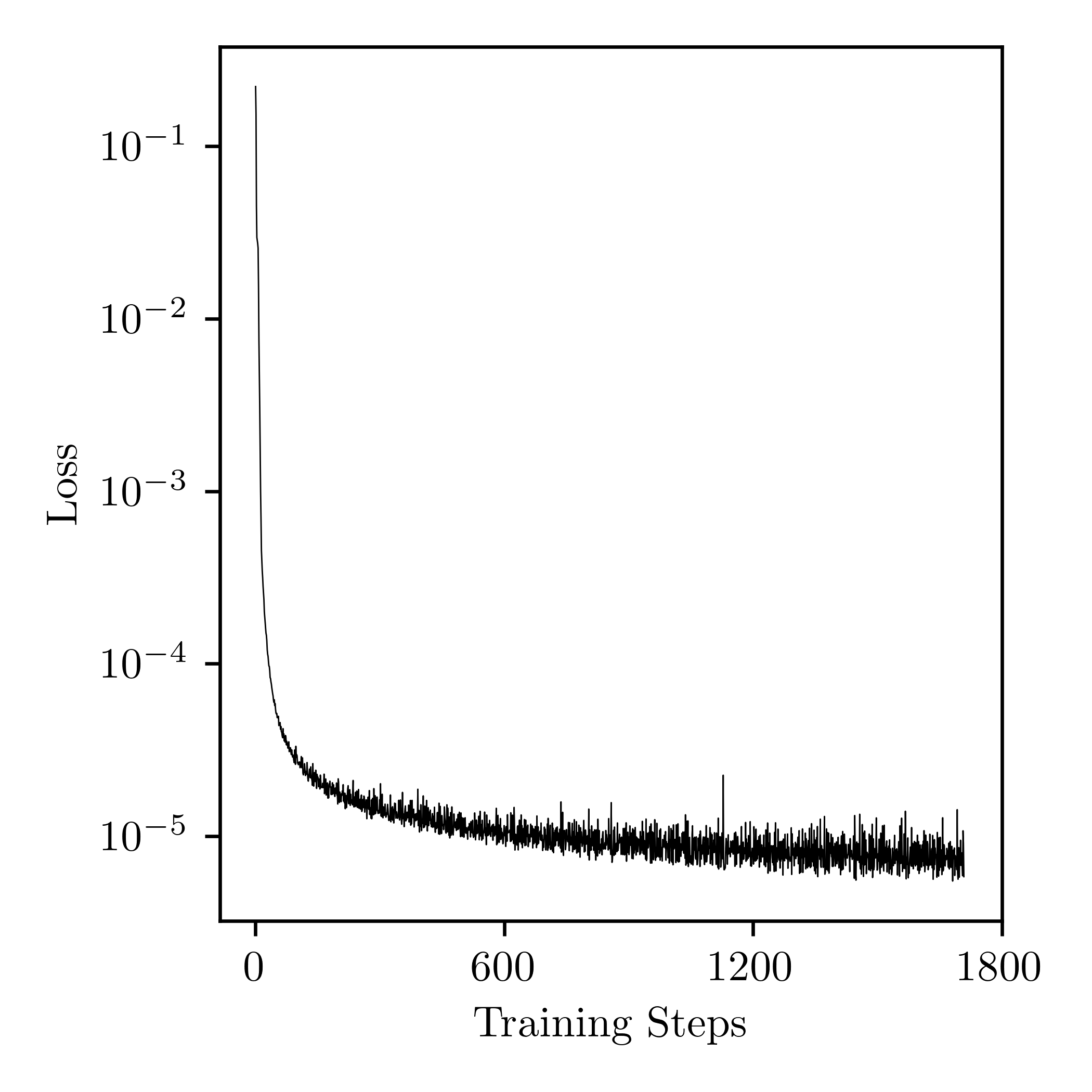}
        \label{subfig:trainloss5}}
    \subfigure[]{%
        \includegraphics[width=3.15in]{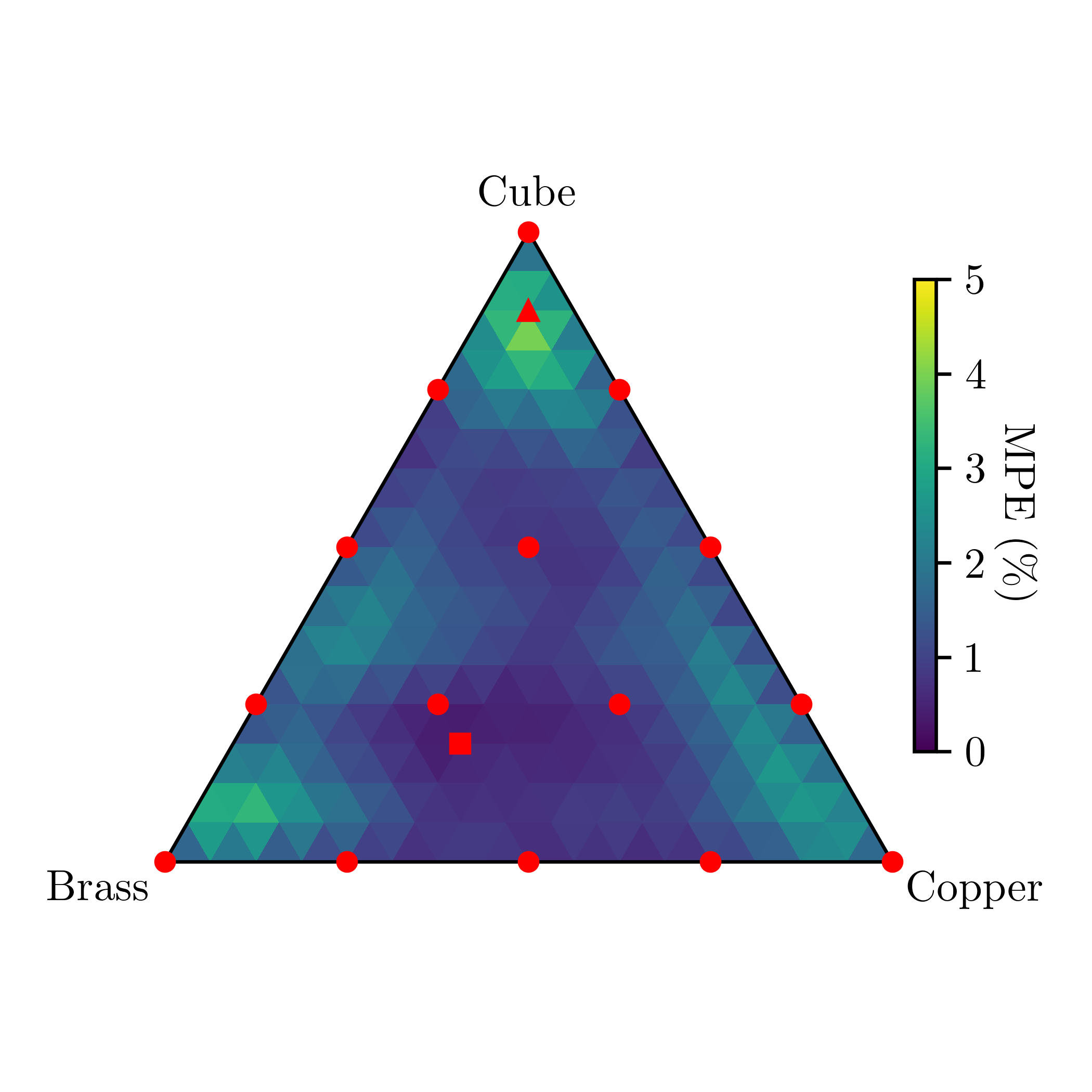}
        \label{subfig:heatmap5}}
    \caption{\subref{subfig:trainloss5} Training loss curve for the pICNN trained with simulations performed with textures marked at the red points in Figure~\ref{fig:textri} (i.e., 15 textures), and \subref{subfig:heatmap5} a ternary heat map displaying MPE between the pICNN and CPFEM predictions at every texture mixture considered in Figure~\ref{fig:textri}, where each element (triangle) is colored based on its nodal point values (i.e., the MPE values for each texture). Circular red points on the plot correspond to points utilized for pICNN training, while the red triangle indicates the texture with the highest MPE, and the red square indicates the texture with the lowest MPE.}
    \label{fig:errors5}
\end{figure}

Here, we present the results from the model trained using the pICNN framework. We first show results for the model trained with only a single spread considered, i.e., $\theta=$\SI{5}{\degree}. We trained the model using the yield curves elucidated at 15 different textures evenly-spaced in texture-weight space (see: red points in Figure~\ref{fig:textri}, corresponding to a 2nd level discretization of the ternary diagram as described in Section~\ref{subsec:texture}). We first present the training loss in Figure~\ref{subfig:trainloss5}. We note that the training loss represents the comparison of {\emph{all grid points}} in stress-space (see: Section~\ref{subsec:picnn} and~\citep{fuhgcnn}), not just the points that represent the yield curve. 

To understand both the goodness-of-fit and applicability of the model, we investigate the error in yield predictions between results calculated from CPFEM simulations and the yield prediction from the pICNN-trained model. We utilize the mean percentage error (MPE) of the predictions {\emph{specifically of the yield points between pICNN and CPFEM predictions}} (as opposed to all grid points in stress-space, as above for the training loss) as these points are the most important for prediction of yield. Further, mean absolute error (a close relative) has been shown~\citep{willmott2005advantages} to be an appropriate error metric for such physics-inspired machine learning with an objective to predict natural phenomena. We use the pICNN model to make yield curve predictions for every texture considered in Figure~\ref{fig:textri}. For each texture, we make predictions specifically along the 13 load vectors at which the CPFEM simulations were performed. We thus calculate the MPE, $e$, for each texture/sample considering the yield results of the CPFEM simulations and the pICNN predictions, or:
\begin{equation}
    \label{eq:mpe}
    e = \frac{\sum_{i=1}^{13}{\left( \frac{\left| r_i^p - r_i^t \right|}{r_i^t} \cdot 100 \right)}}{13}
\end{equation}
where $r^p$ is the magnitude of the load vector at the point of yield for the pICNN prediction and $r^t$ is the magnitude of the load vector at the point of yield for the CPFEM simulation. These values are plotted in Figure~\ref{subfig:heatmap5}, where the triangular elements are colored based on their nodal-point MPE values (the nodal points being the various textures presented in Figure~\ref{fig:textri}). We note that the trained model generates a yield prediction in approximately \SI{2.60}{\second} (on average over 1000 random predictions), an approximately 242-fold reduction compared to a CPFEM simulation runtime of \SI{629}{\second} (on average over 100 random simulations).

%%%%%%%%%%%%%%%%%%%%%
\section{Discussion}
\label{sec:Discussion}

\subsection{Discussion of Error}
\label{subsec:error}

Inspecting first the training loss in Figure~\ref{subfig:trainloss5}, we note that the pICNN is able to both show a low loss value (on the order of \SI{1e-5}{}) as well as saturation behavior (i.e., the training loss is not appreciably decreasing) by approximately 900 training steps. Collectively, this lends us confidence on the training process.

Our greater interest, however, is in the trained model's ability to offer predictions at points beyond the training data. The results of such errors, using an MPE metric, are presented in Figure~\ref{subfig:heatmap5}. We first note that the MPE for each sample is less than 5\%. We present statistics regarding the errors considering all samples, training samples, and test samples in Table~\ref{tab:errors5}. Collectively, these values indicate that the trained model is generally adept at predicting the behavior of samples between training points in this complex texture-weight space. Further, and as expected, we note that the average MPE values are less in the training data than in the test data.
\begin{table}[htbp!]
    \centering
    \begin{tabular}{c c c}
    \hline
    {\bf Samples} & {\bf Mean} & {\bf Standard Deviation} \\
    \hline
    All & 1.34 & 0.80 \\
    Training & 0.72 & 0.13 \\
    Test & 1.41 & 0.81 \\
    \hline
    \end{tabular}
    \caption{Error statistics from the model trained with samples with texture spread $\theta=$~\SI{5}{\degree}.}
    \label{tab:errors5}
\end{table}

\begin{figure}
    \centering
    \subfigure[]{%
	\includegraphics[width=3.15in]{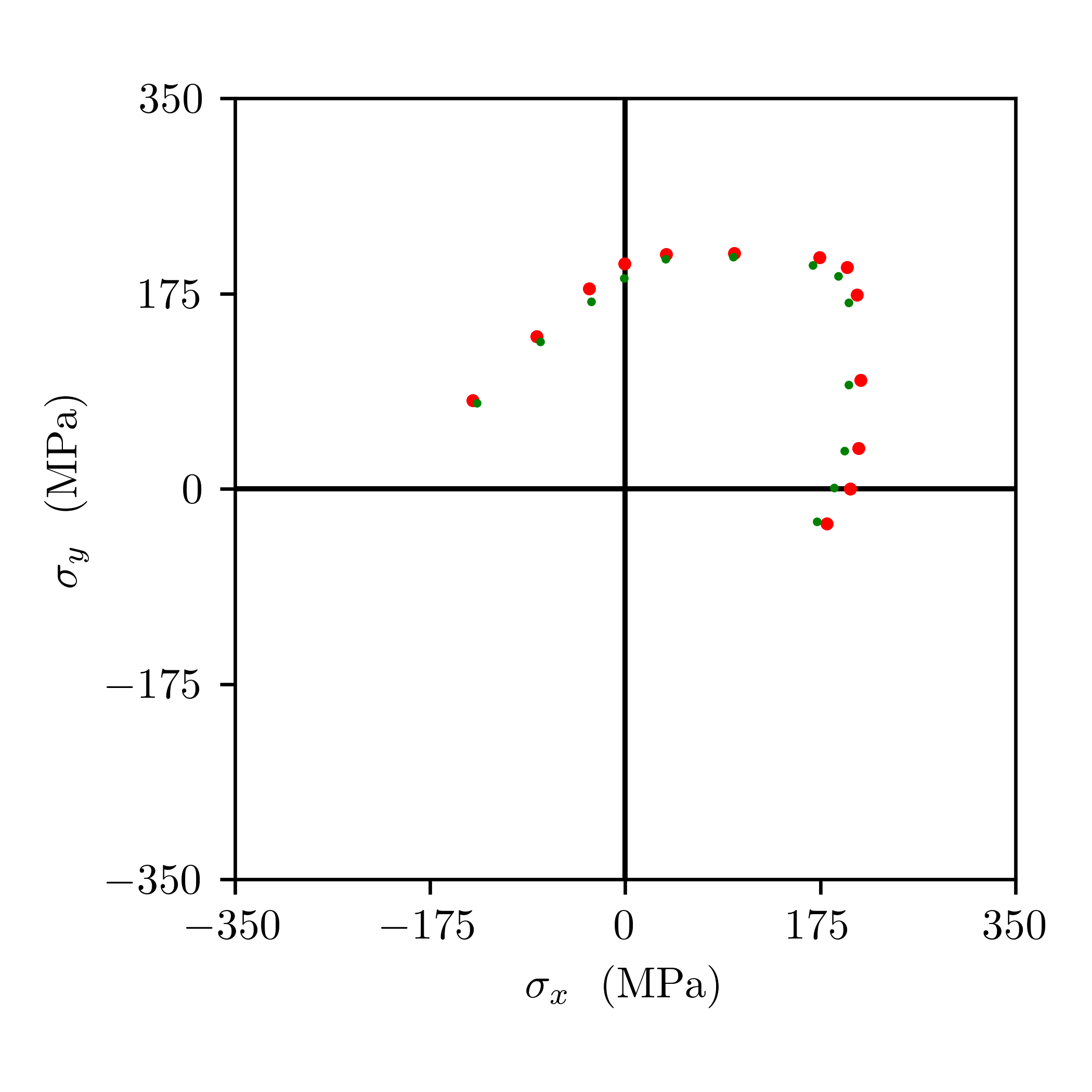}
        \label{subfig:worst5}}
    \subfigure[]{%
	\includegraphics[width=3.15in]{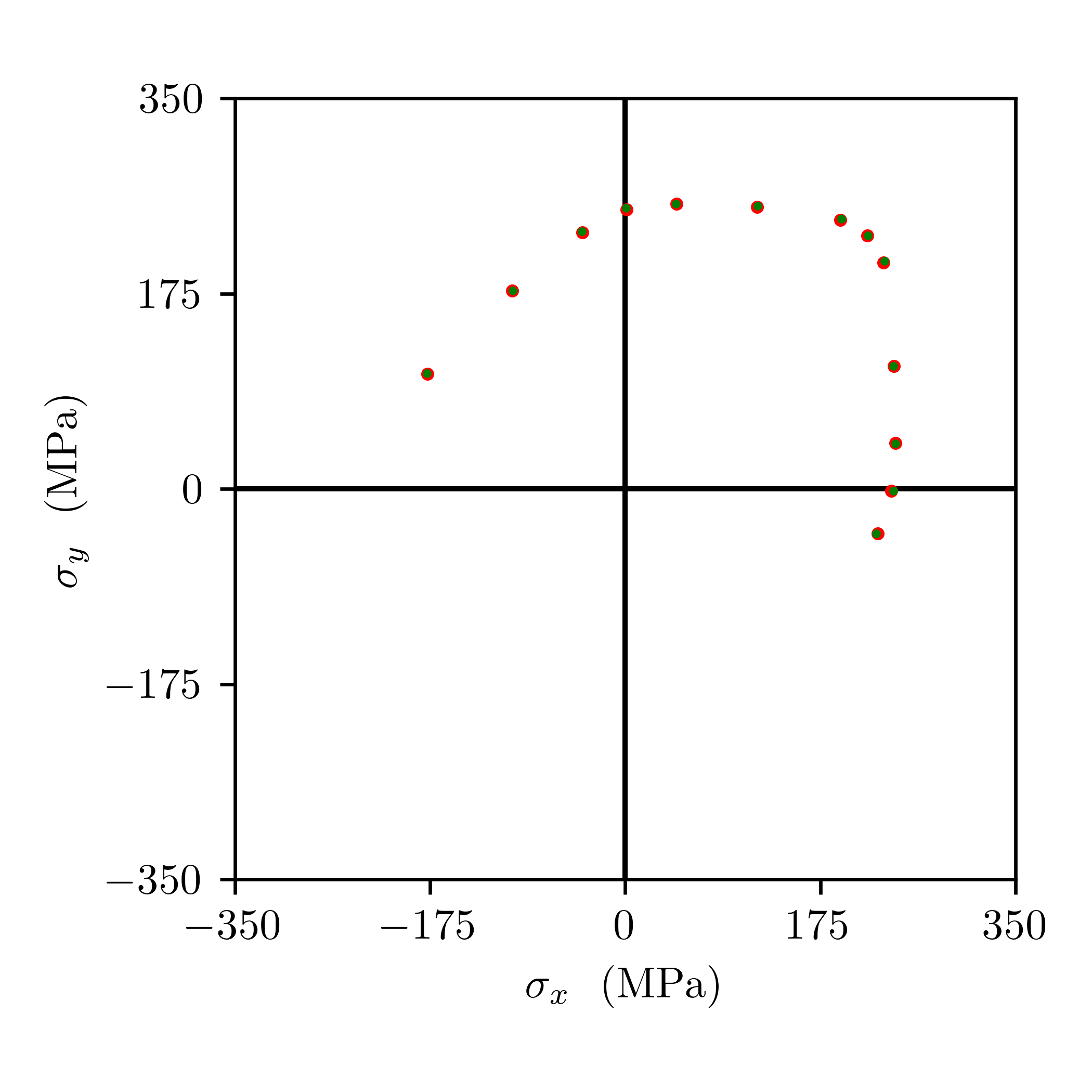}
        \label{subfig:best5}}
    \caption{CPFEM versus pICNN yield predictions for~\subref{subfig:worst5} the sample with the worst MPE, and~\subref{subfig:best5} the sample with the best MPE. Here, red points indicate CPFEM predictions while green points indicate pICNN predictions.}
    \label{fig:worstbest5}
\end{figure}

The worst-case sample in the space we have considered is that with texture weights $(0.8750, 0.0625, 0.0625)$ (ordered $w_c$, $w_b$, $w_{Cu}$), which we find to have an MPE of 4.35\%. Conversely, the best-case sample is that of $(0.1875, 0.5000, 0.3125)$, which displays an MPE of approximately 0.33\% (we note that this sample, interestingly, was not used for training). The specific yield surfaces for these two scenarios (best- and worst-case) are displayed in Figure~\ref{fig:worstbest5}.

Considering all samples, we note that the MPE tends to be lowest near training points. Conversely, we observe that the highest values of MPE are found near the vertices of the texture-weight space (i.e., near the points of unimodal textures). We interpret the higher MPE in these regions to be a direct function of the relative lack of trimodal training points in their vicinity. In other words, despite being relatively close to both bimodal and unimodal training points, these points are far from the nearest trimodal training points and thus have the highest degree of difficulty in producing a trimodal prediction. We thus have indication that the influence of training point densities is inhomogeneous across texture-weight space, and susceptible to lack of training points near ``transitional'' portions of the parameter space considered.

\subsection{Influence of Location of Training Points}
\label{subsec:trainpts}

Based on the discussion above, we wish to gain an understanding of the influence of the location of our training points in texture-weight space on the error in predictions. In the following subsections, we make modifications to the choice of samples used for training and discuss the results. Here, we continue to consider only a single spread of texture, $\theta=$\SI{5}{\degree}, for computational expediency.

\subsubsection{Consequences of Additional Training Points}
\label{subsubsec:addpts}

\begin{figure}
    \centering
    \subfigure[]{%
	\includegraphics[width=2.1in]{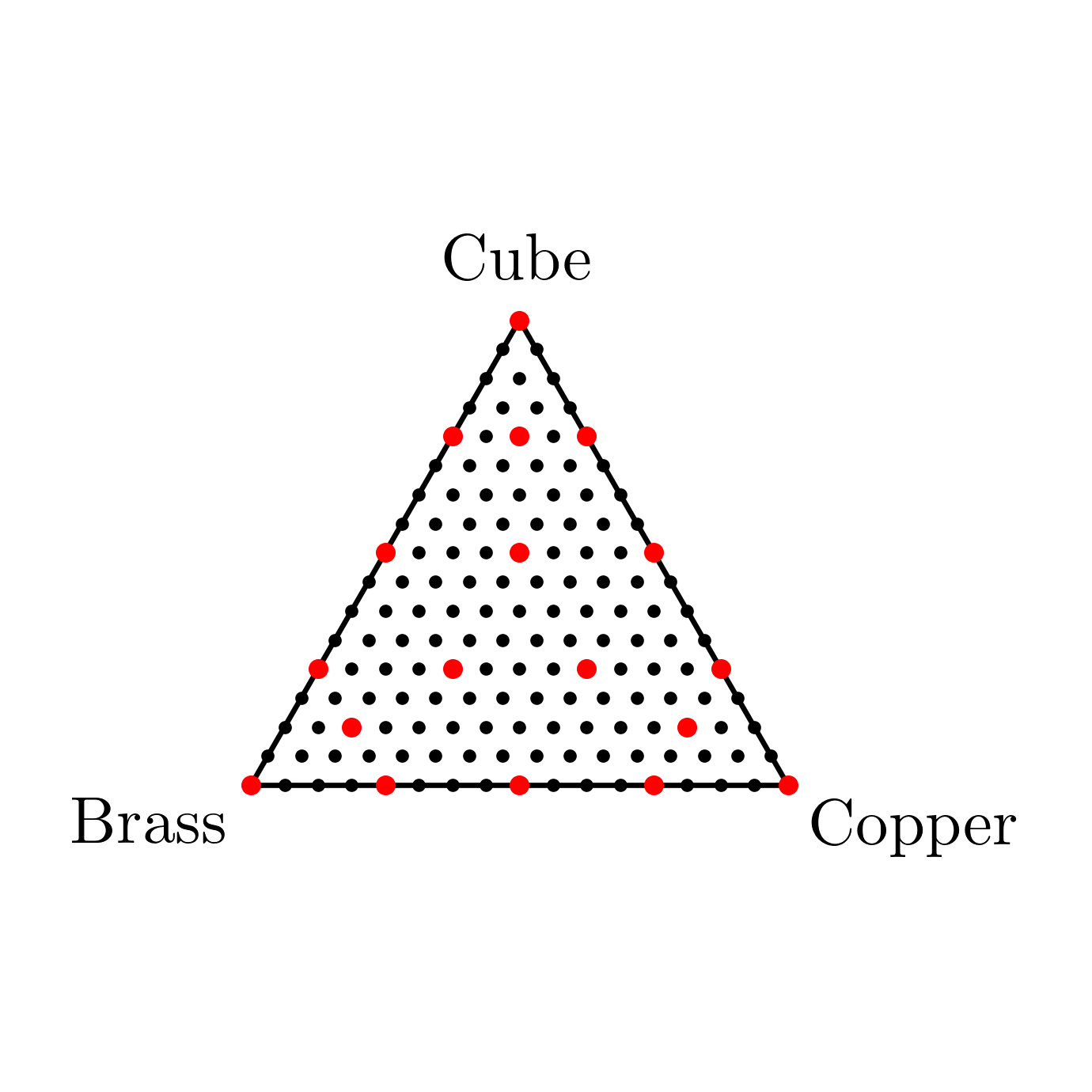}
        \label{subfig:modtrainpts}}
    \subfigure[]{%
	\includegraphics[width=2.1in]{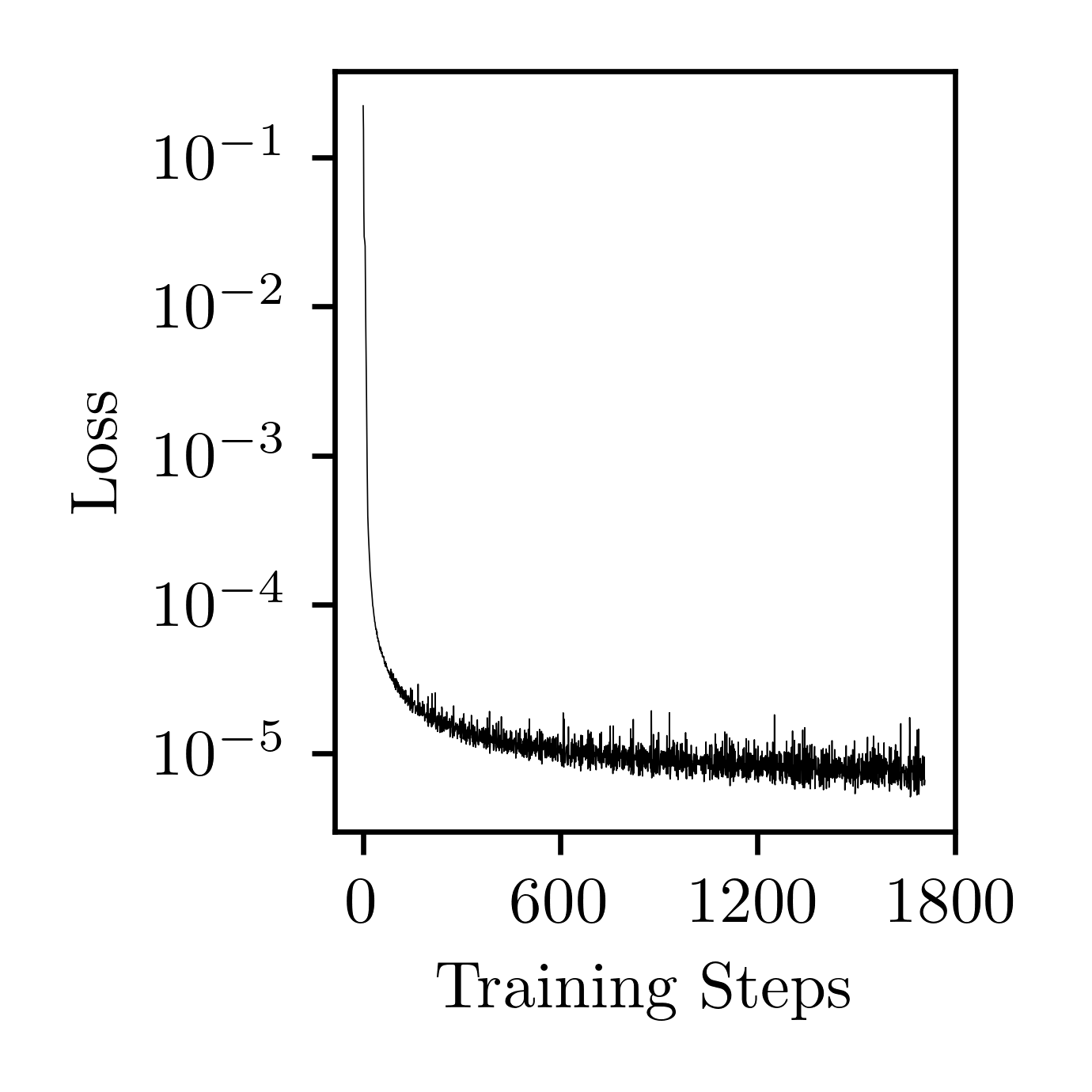}
        \label{subfig:modtrainloss}}
    \subfigure[]{%
	\includegraphics[width=2.1in]{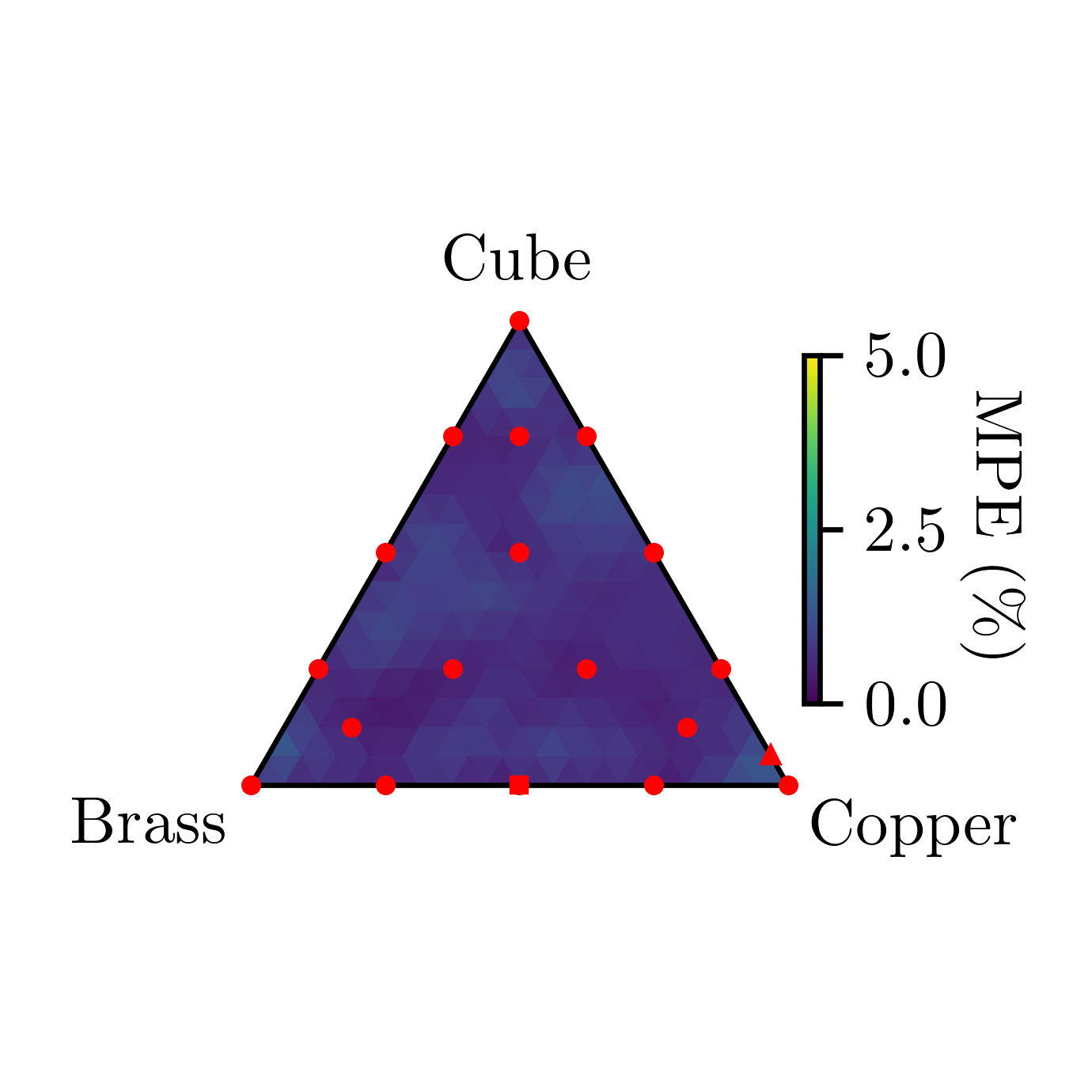}
        \label{subfig:modtrainheat}}
    \caption{\subref{subfig:modtrainpts} Ternary plot depicting a set of modified training points, \subref{subfig:modtrainloss} the training loss, and \subref{subfig:modtrainheat} ternary plot with MPE plotted against each testing sample.}
    \label{fig:modtraining}
\end{figure}

Pursuant to the trends observed in Section~\ref{subsec:error}, we test the effects of providing a higher density of training points near portions of texture-weight space exhibiting the highest values of MPE. Specifically, since the highest MPEs observed in Figure~\ref{subfig:heatmap5} are near the vertices of the ternary heat maps (i.e., the points of unimodal textures), we opt to add 3 training points between the existing trimodal training points and the unimodal training points, as shown in Figure~\ref{subfig:modtrainpts}. We thus train the model with a total of 18 samples, and train to a similar number of training steps as previously. We present the results in Figure~\ref{subfig:modtrainloss} (training loss) and Figure~\ref{subfig:modtrainheat} (ternary plot depicting MPE).

We observe that the model including further training points generally yields better results. We present error statistics considering all samples, training samples, and test samples in Table~\ref{tab:errors5+3}. We find the maximum error to be 1.71\% for the sample $(0.0625, 0.0000, 0.9375)$. The prediction at $(0.8750, 0.0625, 0.0625)$ (previously the worse-case, MPE of 4.35\%) exhibits an MPE of 1.33\% with the addition of the new training points, a threefold reduction from the previous paradigm when considering only 15 training points. Overall, this indicates the advantage of including a higher density of training points near transitional points of the ternary diagram (i.e., areas of transition from unimodal to multimodal textures). We note, however, that even when considering only 15 training points as above, the MPE is generally below 5\%. Thus, such refinement of training points may only be necessary if high accuracy is desired.
\begin{table}[htbp!]
    \centering
    \begin{tabular}{c c c}
    \hline
    {\bf Samples} & {\bf Mean} & {\bf Standard Deviation} \\
    \hline
    All & 0.72 & 0.26 \\
    Training & 0.50 & 0.10 \\
    Test & 0.75 & 0.26 \\
    \hline
    \end{tabular}
    \caption{Error statistics from the model trained with samples with texture spread $\theta=$~\SI{5}{\degree}, with additional samples in the training set.}
    \label{tab:errors5+3}
\end{table}

\subsubsection{Consequences of Random Training Points}
\label{subsubsec:randpts}

\begin{figure}
    \centering
    \subfigure[]{%
	\includegraphics[width=2.1in]{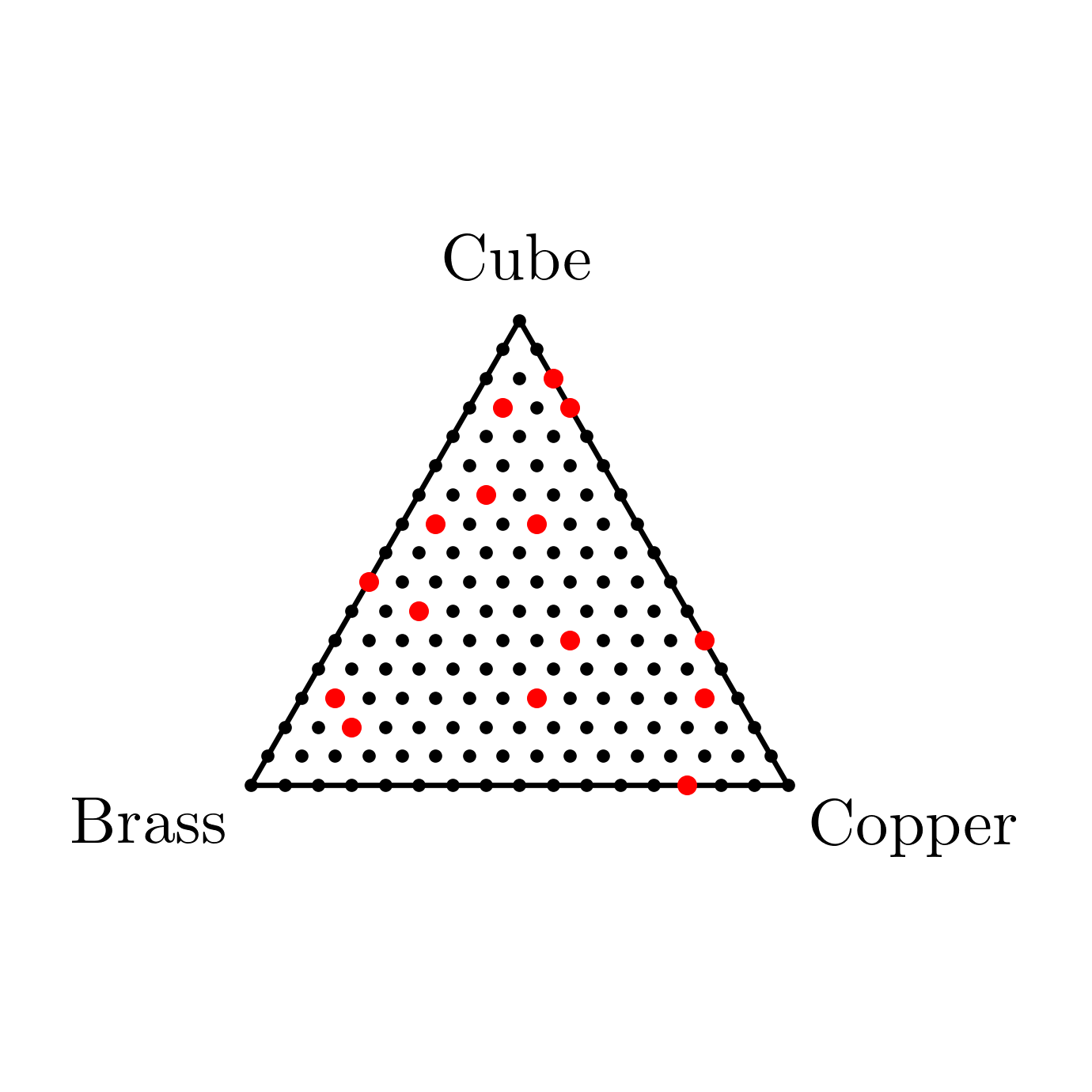}
        \label{subfig:randtrainpts1}}
    \subfigure[]{%
	\includegraphics[width=2.1in]{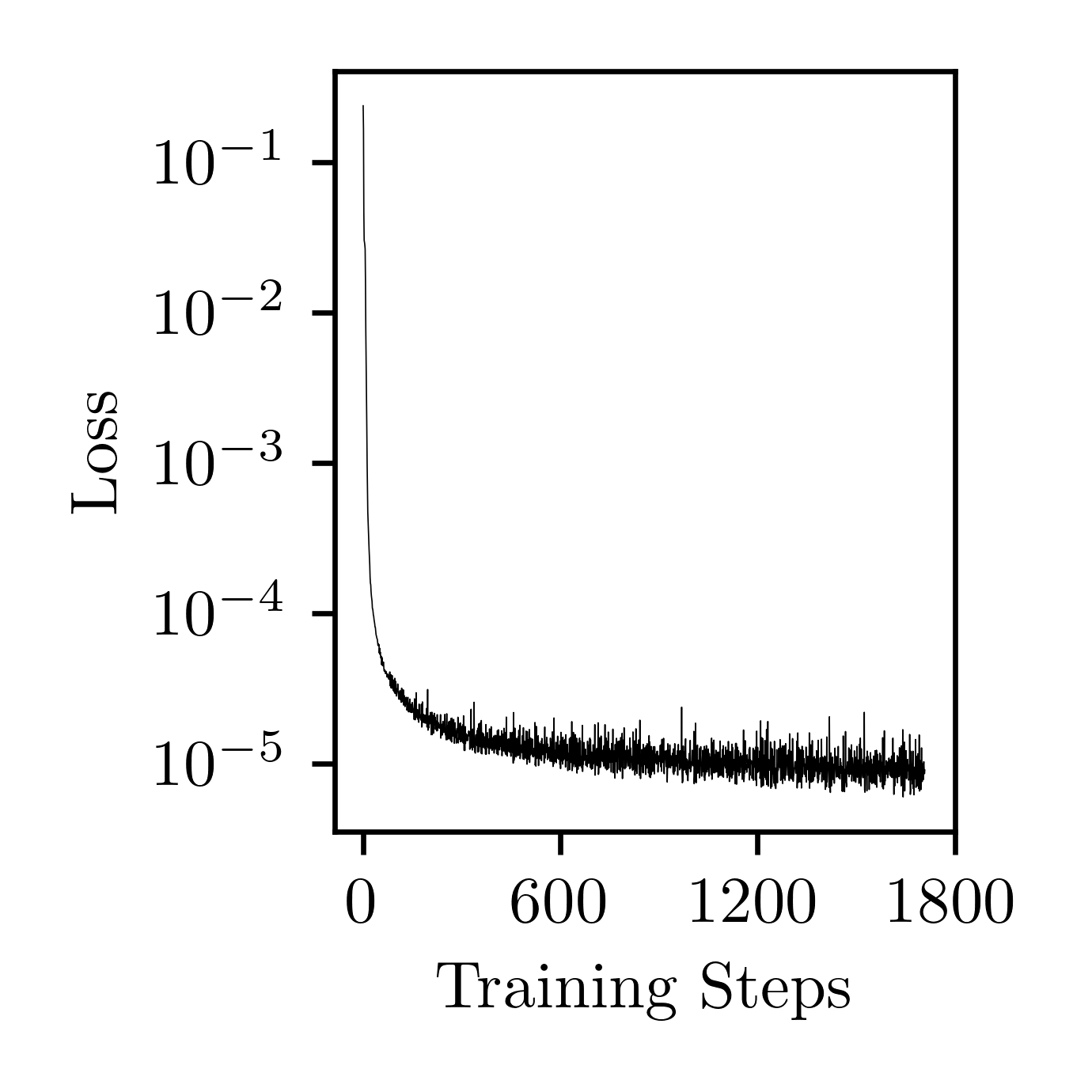}
        \label{subfig:randtrainloss1}}
    \subfigure[]{%
	\includegraphics[width=2.1in]{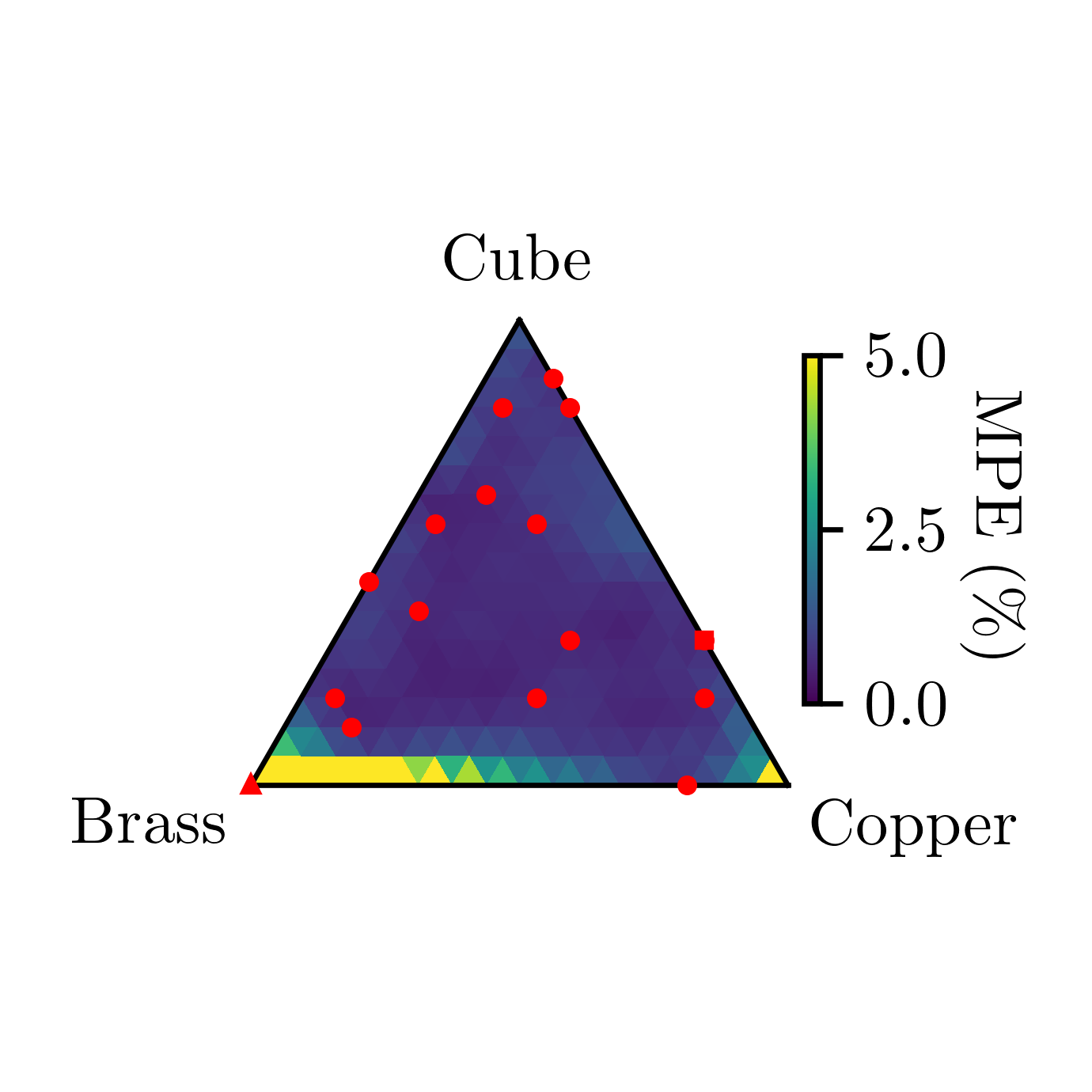}
        \label{subfig:randtrainheat1}}    
    \subfigure[]{%
	\includegraphics[width=2.1in]{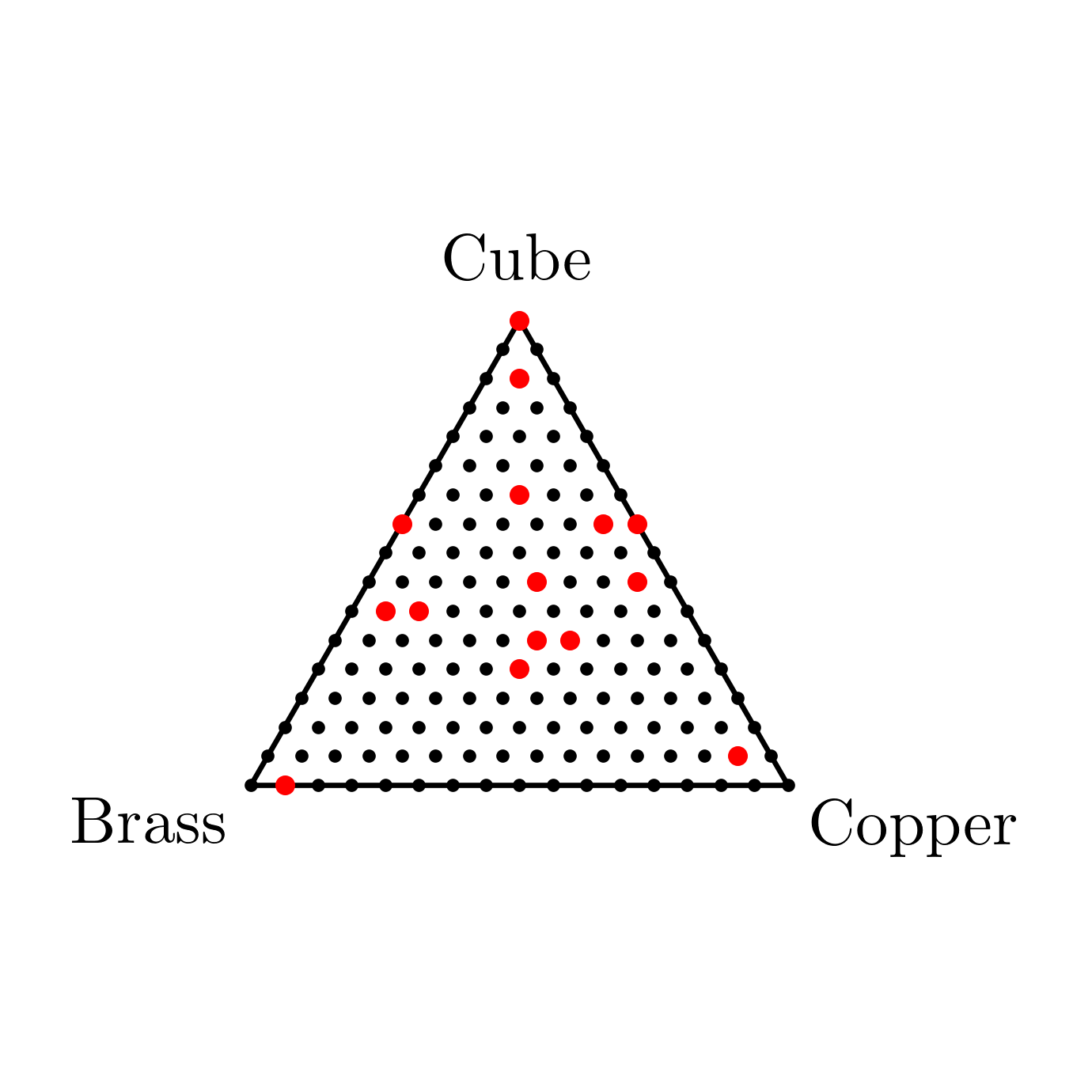}
        \label{subfig:randtrainpts2}}
    \subfigure[]{%
	\includegraphics[width=2.1in]{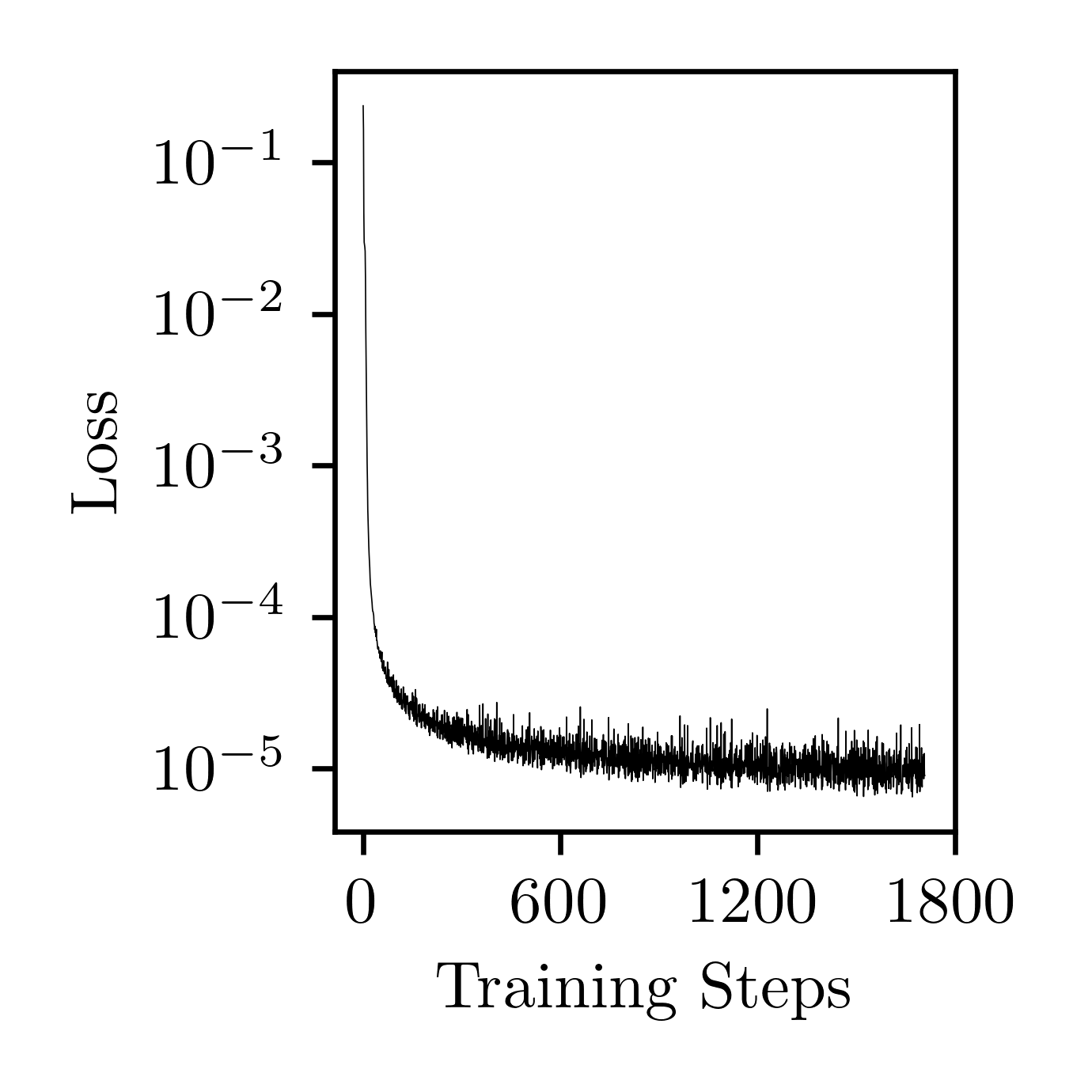}
        \label{subfig:randtrainloss2}}
    \subfigure[]{%
	\includegraphics[width=2.1in]{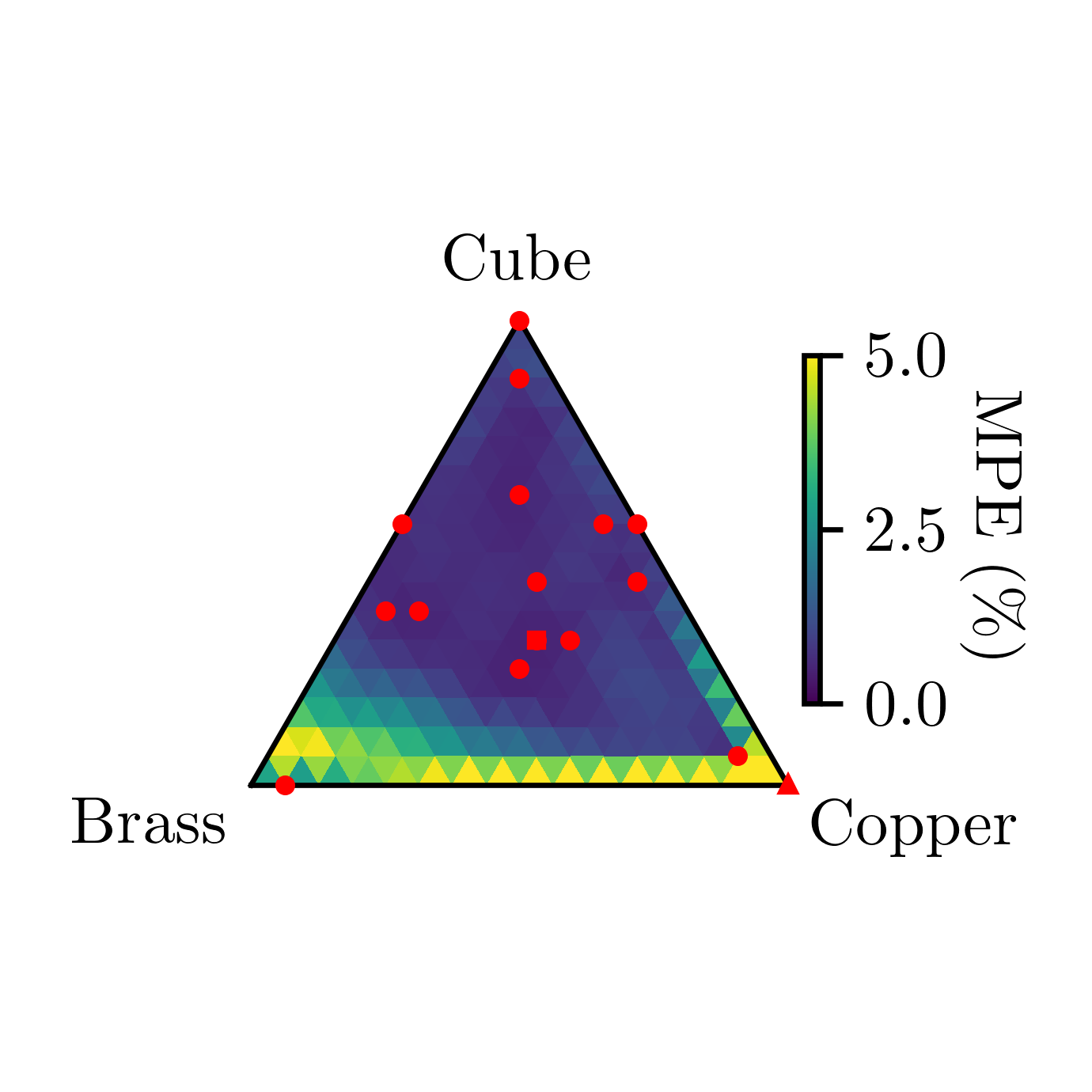}
        \label{subfig:randtrainheat2}}   
    \subfigure[]{%
	\includegraphics[width=2.1in]{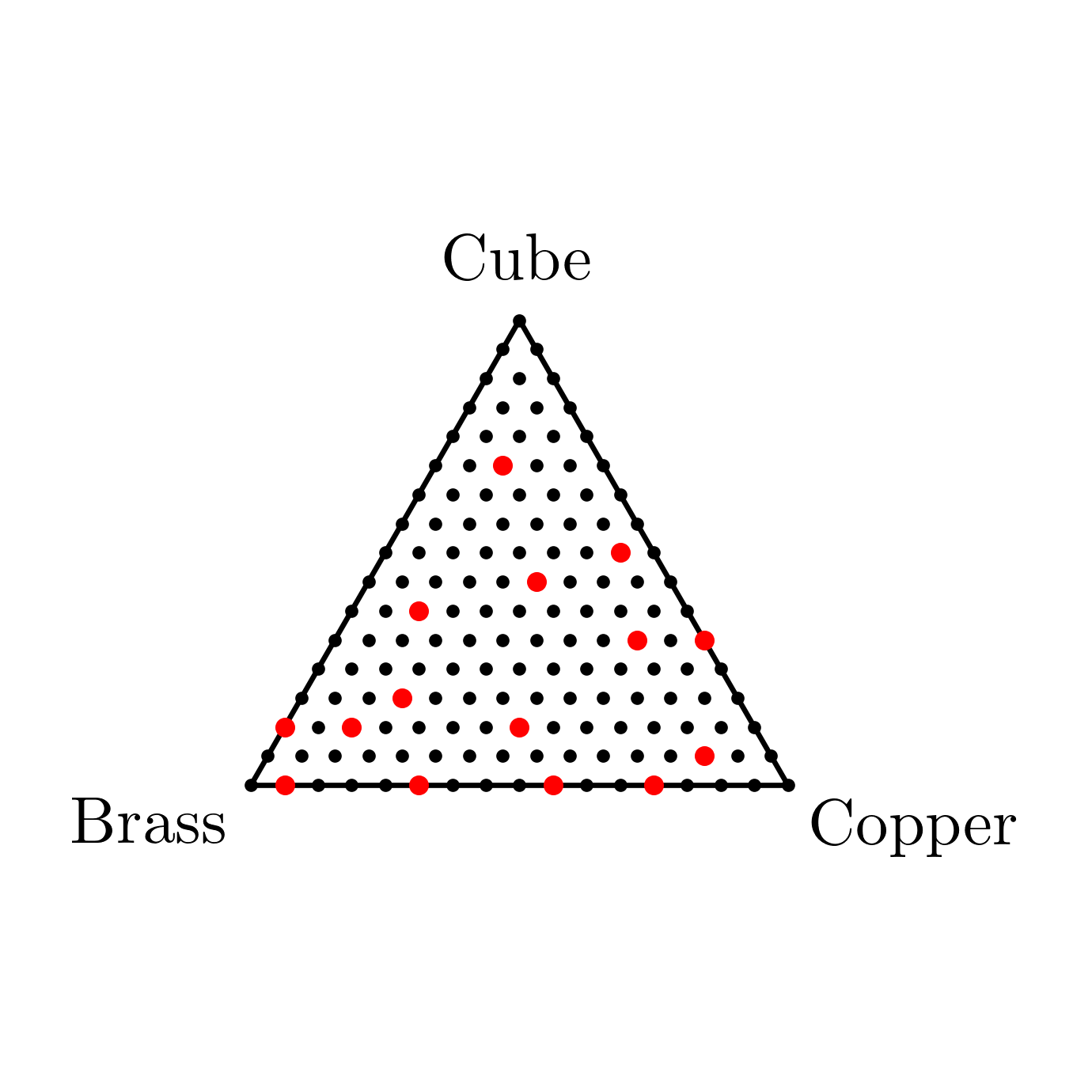}
        \label{subfig:randtrainpts3}}
    \subfigure[]{%
	\includegraphics[width=2.1in]{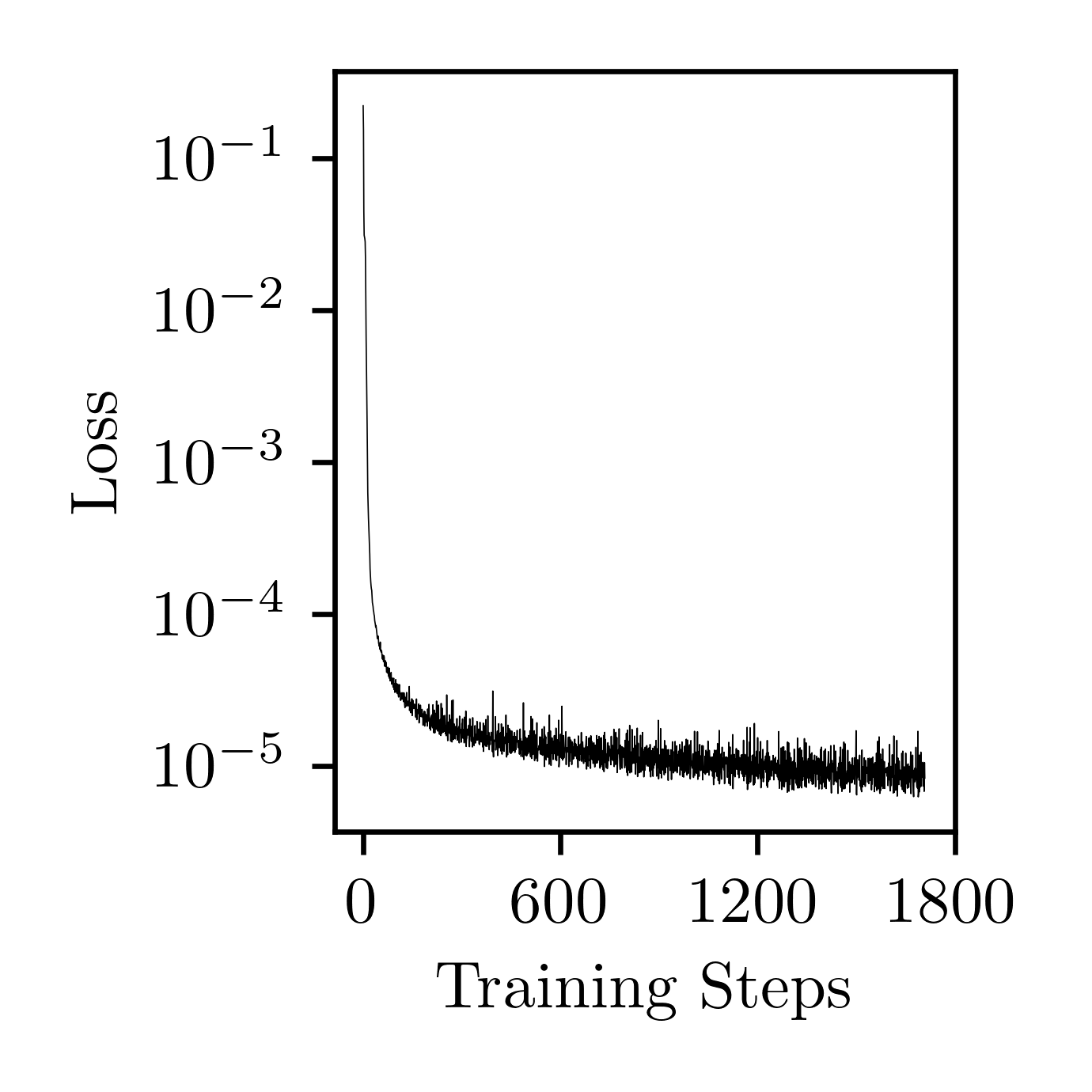}
        \label{subfig:randtrainloss3}}    
    \subfigure[]{%
	\includegraphics[width=2.1in]{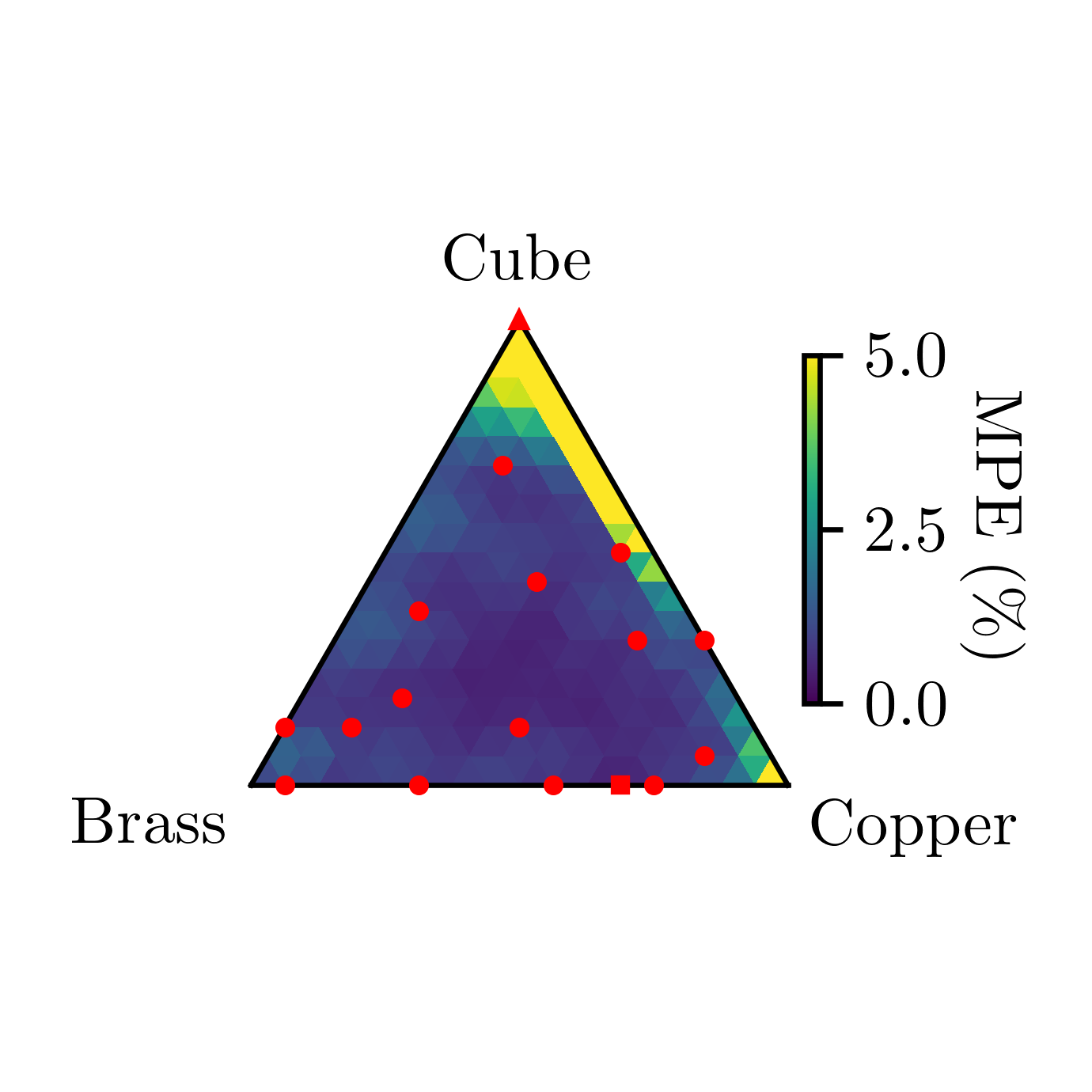}
        \label{subfig:randtrainheat3}}
    \caption{\subref{subfig:randtrainpts1},~\subref{subfig:randtrainpts2},~and~\subref{subfig:randtrainpts3} Ternary plots depicting three sets of random training points, \subref{subfig:randtrainloss1},~\subref{subfig:randtrainloss2},~and~\subref{subfig:randtrainloss3} the training loss, and \subref{subfig:randtrainheat1},~\subref{subfig:randtrainheat2},~and~\subref{subfig:randtrainheat3} ternary plots with MPE plotted against each testing sample.}
    \label{fig:randtraining}
\end{figure}

We next wish to test whether results are sensitive to the careful choice of training points. Previously, we have primarily considered the choice of 15 textures/samples, with careful placement on the ternary diagram corresponding to various unimodal, bimodal, and trimodal textures. Here, we instead choose 15 random samples from the total set of 153. We choose three random sets of 15 samples, as depicted in Figures~\ref{subfig:randtrainpts1},~\subref{subfig:randtrainpts2},~and~\subref{subfig:randtrainpts3}, taking no care to span the parameter space in any purposeful way. We train to a similar number of training steps as previously, and depict the results of the training in Figures~\ref{subfig:randtrainloss1},~\subref{subfig:randtrainloss2},~and~\subref{subfig:randtrainloss3} (training losses) and Figures~\ref{subfig:randtrainheat1},~\subref{subfig:randtrainheat2}~,~and~\subref{subfig:randtrainheat3} (ternary plots depicting MPE).

We note first the acceptable training loss in all three cases, indicating proper pICNN functionality. However, when inspecting the ternary MPE heat maps, we observe the overall distribution of MPE is worse than when the training points are purposefully placed. We present error statistics considering all samples (training and test) for the three models in Table~\ref{tab:errorsrand}. We further calculate the maximum values of MPE as 53.71\%, 38.36\%, and 56.16\%. All of these values are higher than those presented in Section~\ref{subsec:error}---indeed an order of magnitude greater. Generally, MPE values are (unsurprisingly) highest in regions of texture-weight space containing fewer training points, and lower in regions containing a higher density of training points. This is especially acute when missing unimodal or bimodal training points. We interpret this behavior to be a consequence of the pICNN framework's complete lack of training in these regimes, and thus inability to properly make predictions in these directions, similar to the above-discussed phenomena of behavior near the ``transitional'' portions of the ternary diagram. Overall, we find that the careful placement of training points is key in making acceptable predictions over texture-weight space, as random placement leads to unacceptably high errors due to its poor coverage of the various texture mixtures (i.e., unimodal, bimodal, trimodal) necessary to generate a comprehensive model. Together with the previously presented results, we interpret this to indicate that the yield relationship is a highly non-linear function of the texture descriptors, especially near the transitional portions of the texture-weight space.
\begin{table}[htbp!]
    \centering
    \begin{tabular}{c c c}
    \hline
    {\bf Training Set} & {\bf Mean} & {\bf Standard Deviation} \\
    \hline
    Random 1 & 2.15 & 6.11 \\
    Random 2 & 2.13 & 3.75 \\
    Random 3 & 2.86 & 7.44 \\
    \hline
    \end{tabular}
    \caption{Error statistics considering all samples (test and training) from the model trained with samples with texture spread $\theta=$~\SI{5}{\degree}, utilizing randomly-selected training points.}
    \label{tab:errorsrand}
\end{table}

\subsection{Influence of \texorpdfstring{$\theta$}{theta}}
\label{subsec:theta}

\begin{figure}
    \centering
    \subfigure[]{%
        \includegraphics[width=3.15in]{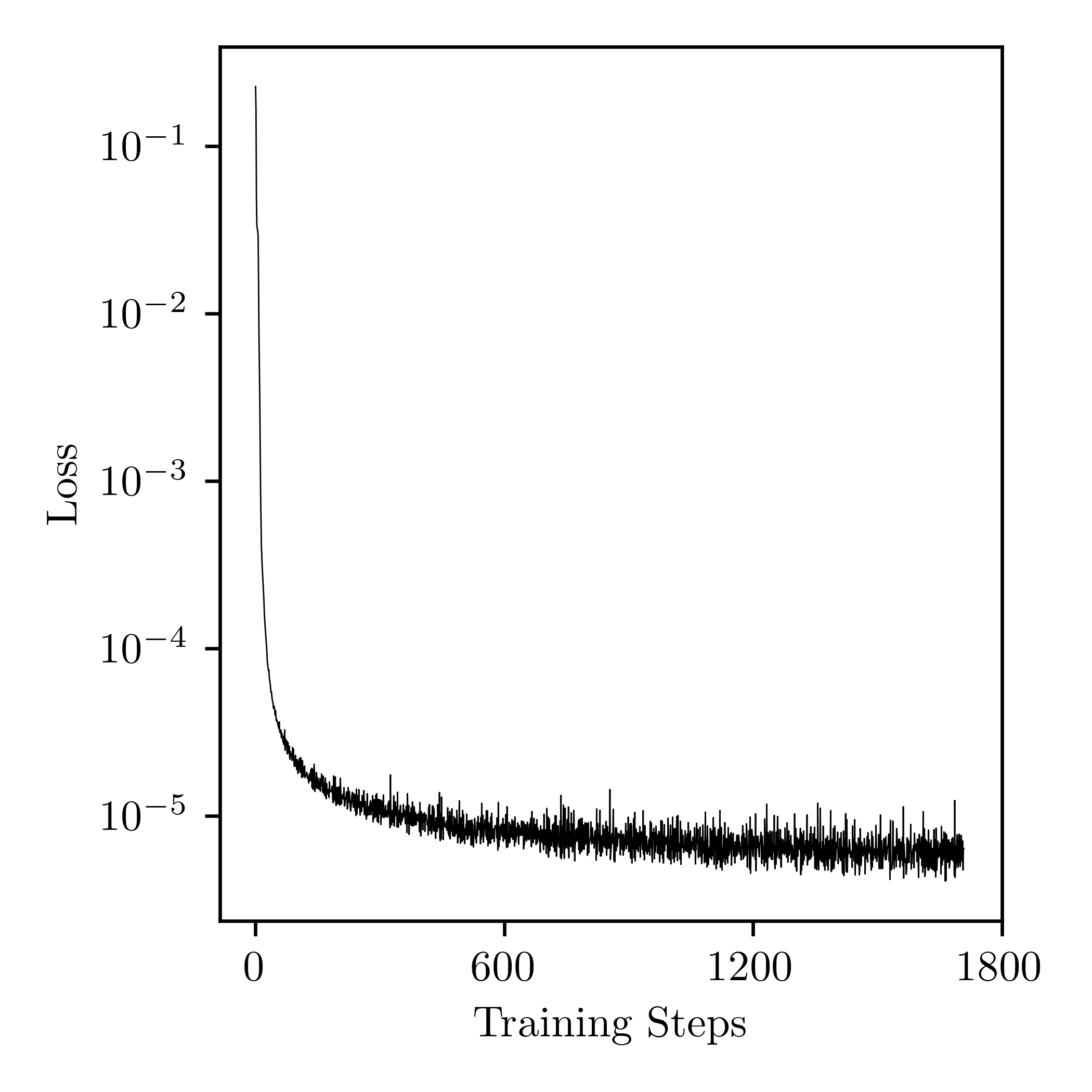}
        \label{subfig:trainloss25}}
    \subfigure[]{%
        \includegraphics[width=3.15in]{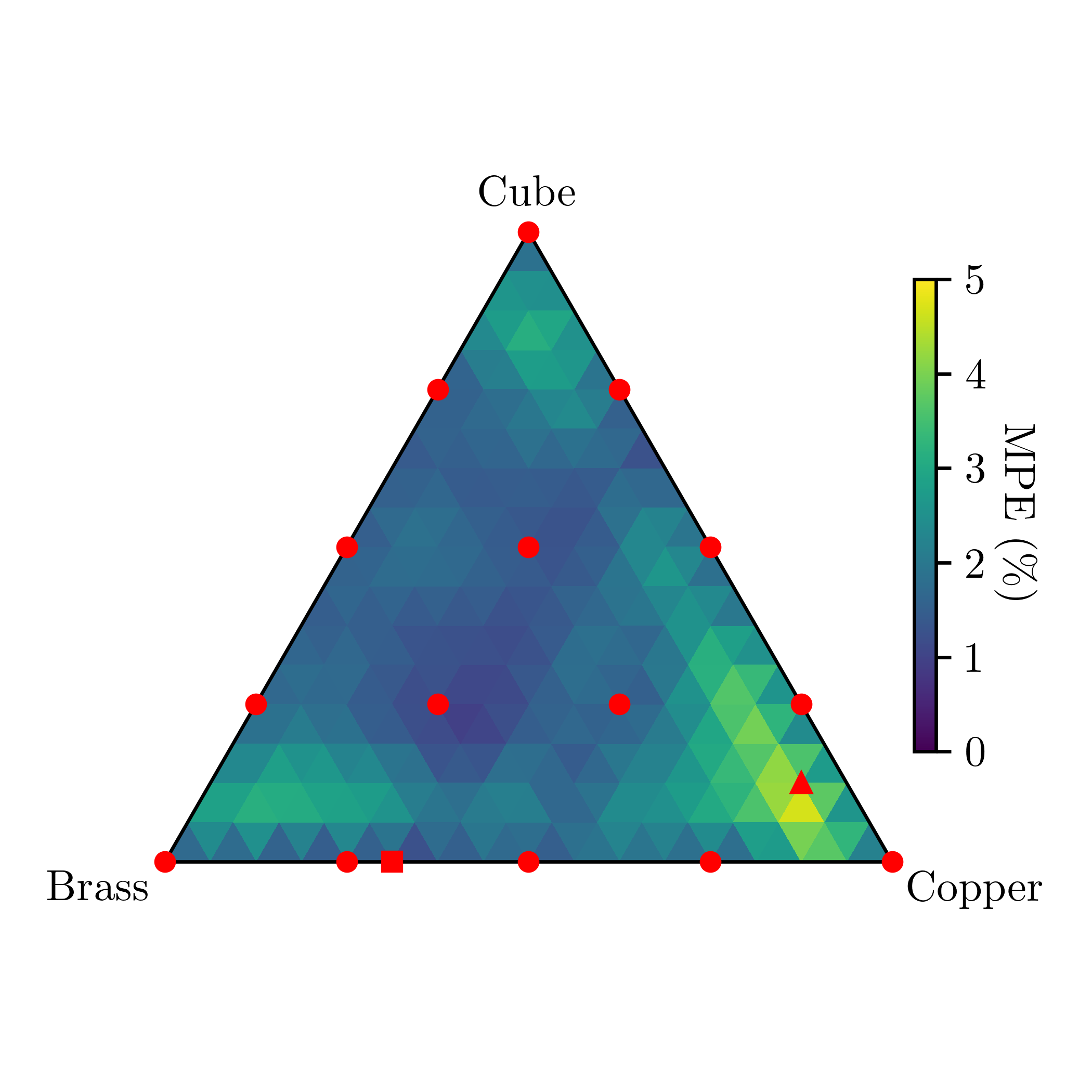}
        \label{subfig:heatmap25}}
    \caption{\subref{subfig:trainloss25} Training loss curve for the pICNN trained with simulations performed with textures marked at the red points in Figure~\ref{fig:textri} (i.e., 15 textures) for samples with $\theta=$~\SI{25}{\degree}, and \subref{subfig:heatmap25} a ternary heat map displaying MPE between the pICNN and CPFEM predictions at every texture mixture considered in Figure~\ref{fig:textri}, where each element (triangle) is colored based on its nodal point values (i.e., the MPE values for each texture). Circular red points on the plot correspond to points utilized for pICNN training, while the red triangle indicates the texture with the highest MPE, and the red square indicates the texture with the lowest MPE.}
    \label{fig:errors25}
\end{figure}

Previously, we observed that small values of $\theta$ (i.e, sharper textures) tend to lead to highly-faceted yield surfaces, while larger values of $\theta$ (i.e., more diffuse textures) tend to lead to highly-curved yield surfaces (discussed in-depth in~\citep{fuhgcnn}, theoretical description in~\citep{Kocks2000,backofen}). We also note that the results of~\citep{fuhgcnn} demonstrate the pICNN's capabilities to generate acceptable interpolative (i.e., in-sample) and extrapolative (i.e., out-of-sample) predictions when varying $\theta$.  Here, we vary the spread of the texture modes, $\theta$, to determine the pICNN's performance in this regard when considering more complex material states. We keep the value of $\theta$ equal for all three texture modes considered. Specifically, we additionally consider only $\theta=$~\SI{25}{\degree} to decrease computational cost, again bolstered by the previously published results demonstrating framework capability when varying $\theta$, allowing us to explore a limited space. We train the pICNN using the same training points as shown on the ternary diagram in Figure~\ref{fig:textri}, though now with a different value of $\theta$. We train to a similar number of training steps as previously. We present the training loss in Figure~\ref{subfig:trainloss25}, as well as a ternary heatmap of MPE values in Figure~\ref{subfig:heatmap25}.

\begin{figure}
    \centering
    \subfigure[]{%
	\includegraphics[width=3.15in]{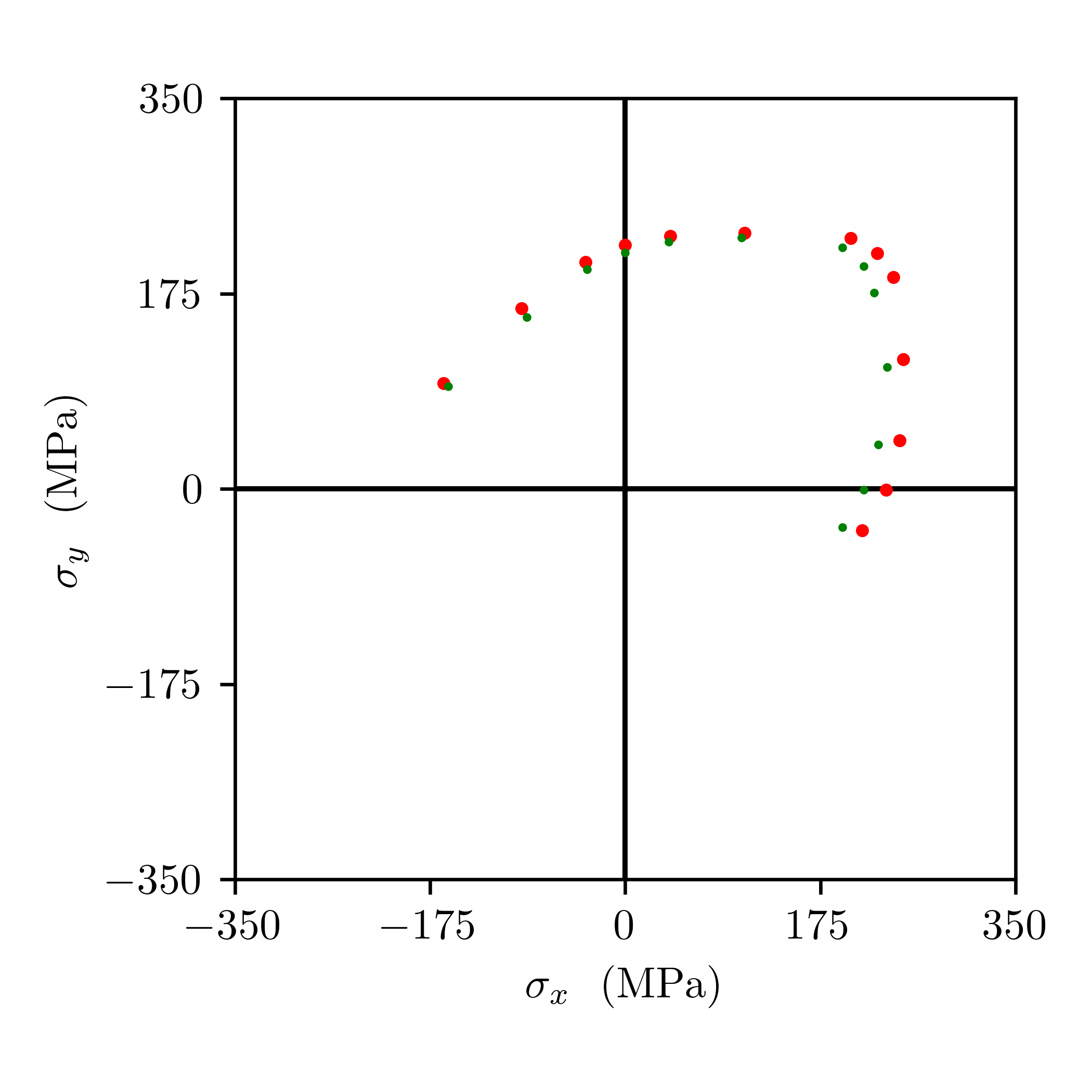}
        \label{subfig:worst25}}
    \subfigure[]{%
	\includegraphics[width=3.15in]{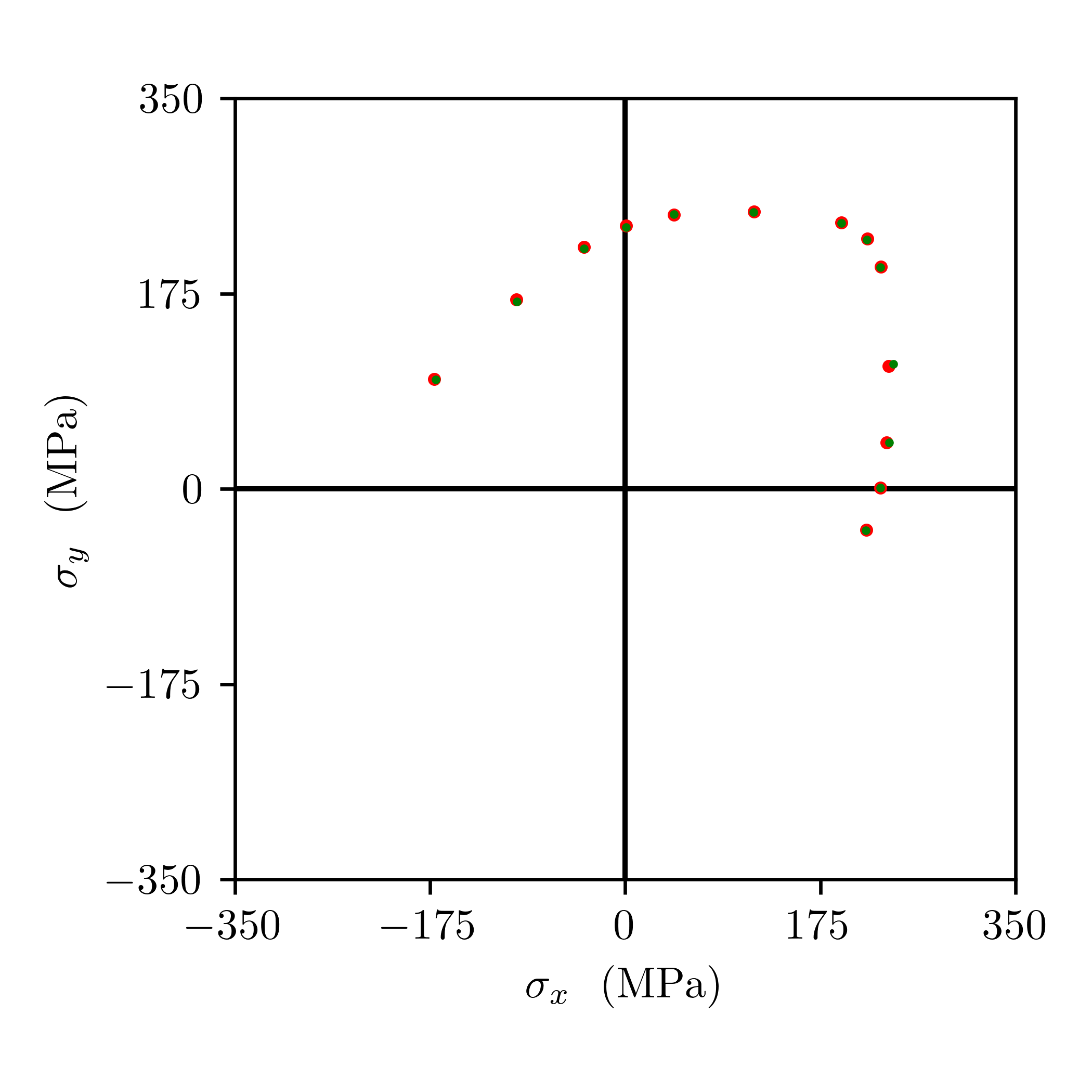}
        \label{subfig:best25}}
    \caption{CPFEM versus pICNN yield predictions for~\subref{subfig:worst5} the sample with the worst MPE, and~\subref{subfig:best5} the sample with the best MPE when considering a texture spread of $\theta=$~\SI{25}{\degree}. Here, red points indicate CPFEM predictions while green points indicate pICNN predictions.}
    \label{fig:worstbest25}
\end{figure}

Generally, we see observe higher errors here than when considering samples with $\theta=$~\SI{5}{\degree}. To quantify this, we present error statistics when considering all samples, training samples, and test samples in Table~\ref{tab:errors25}. Further, we note that the best case scenario is that of $(0.0000, 0.6875, 0.3125)$, which displays an error of 0.51\%, and the worst case scenario is that of $(0.1250, 0.0625, 0.8125)$, which displays an error of 5.02\%. Further, we present the yield surfaces for the best and worst case scenarios in Figure~\ref{fig:worstbest25}. Overall, we again highlight the generally acceptable degree of error, which lends confidence in the framework's ability to learn behavior with various strengths of texture.
\begin{table}[htbp!]
    \centering
    \begin{tabular}{c c c}
    \hline
    {\bf Samples} & {\bf Mean} & {\bf Standard Deviation} \\
    \hline
    All & 1.94 & 0.82 \\
    Training & 1.31 & 0.12 \\
    Test & 2.01 & 0.83 \\
    \hline
    \end{tabular}
    \caption{Error statistics from the model trained with samples with texture spread $\theta=$~\SI{25}{\degree}.}
    \label{tab:errors25}
\end{table}

\subsection{Randomization of Error Estimation}
\label{subsec:randerror}

While the estimation of error presented in the previous section provides a detailed understanding of goodness-of-fit in the parameter space considered (i.e., across the ternary diagram(s)), it is computationally intensive to perform a large number of CPFEM simulations to fill these spaces. We anticipate that such an error estimation paradigm will become increasingly intractable as the dimensionality of the problem is increased (i.e., as more material descriptors are considered during training and prediction). We thus find it advantageous to find a more efficient way to quantify error and qualify the model as appropriately equipped.

\begin{figure}
    \centering
    \subfigure[]{%
	\includegraphics[width=3.15in]{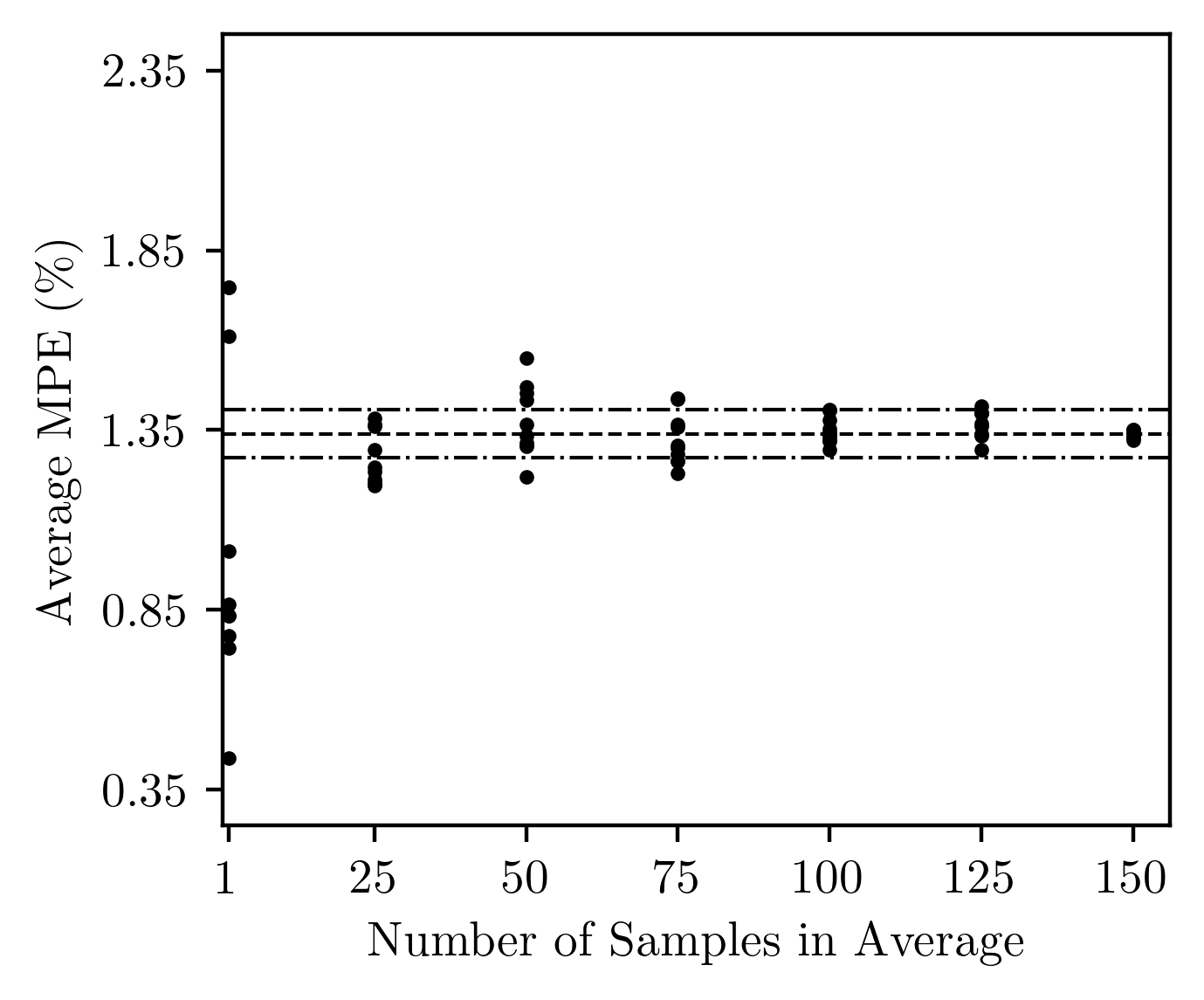}
        \label{subfig:randerrorsamp}}
    \subfigure[]{%
	\includegraphics[width=3.15in]{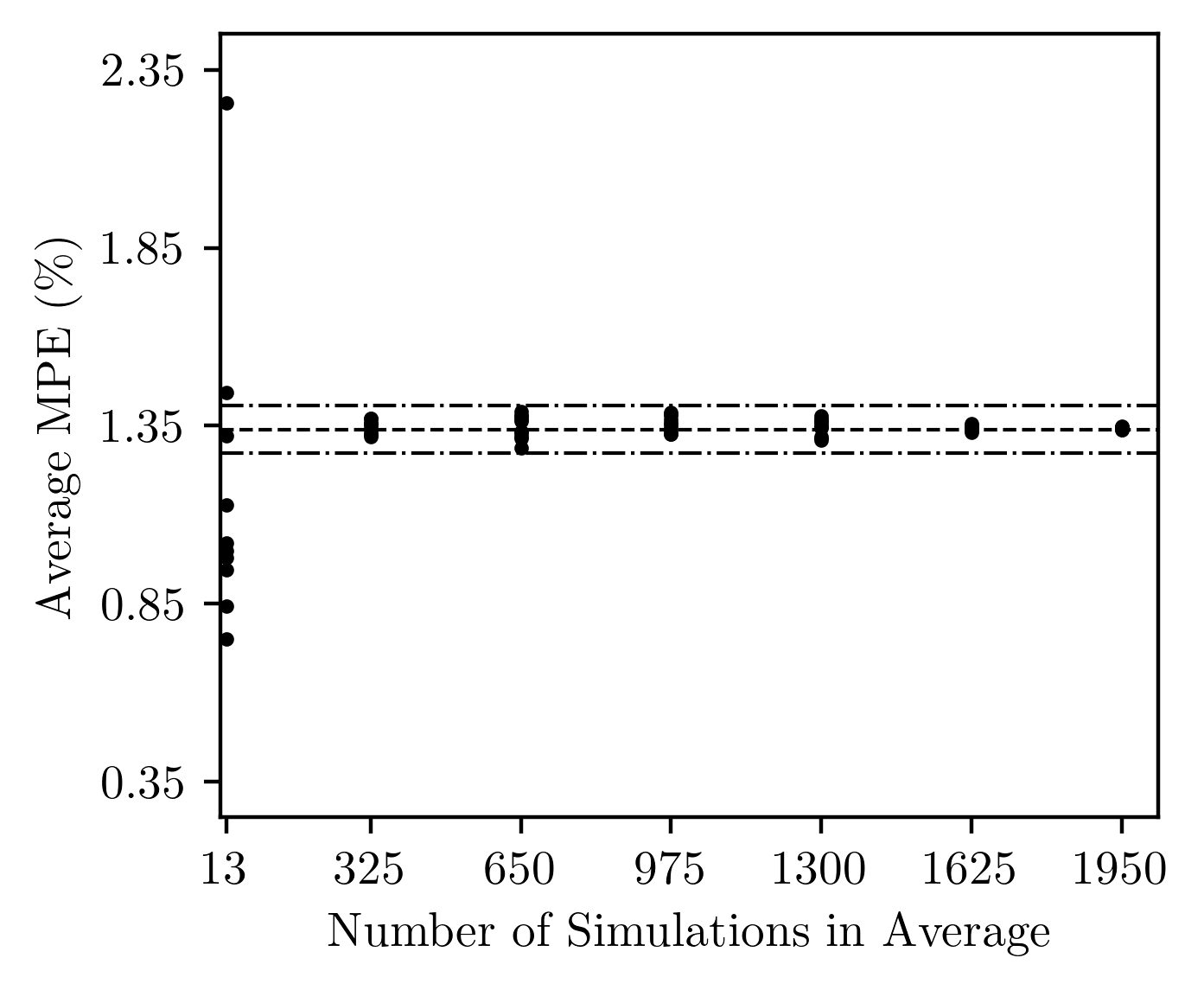}
        \label{subfig:randerrorsims}}
    \caption{Calculation of global average MPE considering the calculation over~\subref{subfig:randerrorsamp} a variable number of samples, including all 13 simulations for each sample considered (using 10 random sets of samples for each number of samples considered) and~\subref{subfig:randerrorsims} a variable number of simulations (using 10 random sets of simulations for each number of simulations considered). The dashed line in both plots indicates the global error when considering all samples/simulations collectively, and the dot-dashed lines indicate $\pm5\%$ of the total average.}
    \label{fig:randerror}
\end{figure}

Consequently, we here explore a reduced-order error calculation. For demonstration, we perform this study considering simulations with $\theta=$~\SI{5}{\degree}. Here, we first vary the number of samples considered for the calculation of error. For a given number of samples considered, we randomly select that number of samples and calculate the MPE between the pICNN predictions and the CPFEM predictions for the simulations performed on these samples alone (i.e., considering the 13 simulations per sample). We find the mean MPE among these samples. We perform this 10 times for each number of samples considered, using a different random set of samples for each number of samples considered. We plot the results in Figure~\ref{subfig:randerrorsamp}. We note that when the error calculation considers a smaller number of samples, the results tend to be more varied. As we consider an increasingly larger number of samples, the error estimations tend to become more consistent, and indeed converge to the error considering all samples (i.e., $1.34\%$, see: Section~\ref{subsec:error}). We observe that error calculations are within 5\% of the converged value when considering approximately 100 or more samples (or 1,300 simulations).

Similarly, we next consider an error metric by varying the number of simulations considered for the calculation of error. For a given number of simulations considered, we choose a random set of simulations (which may or may not come from the same sample) and calculate the MPE between the pICNN predictions and the CPFEM predictions for these simulations alone. We note that we perform this analysis at the same number of simulations that are considered in the previous analysis considering errors calculated on a per-sample basis (though again, we do not restrict that the simulations must come from the same sample). We perform this 10 times for each number of simulations considered, using a different random set of simulations for each number of simulations considered. We plot the results in Figure~\ref{subfig:randerrorsims}. We note that similar trends as described above persist here, though the error estimations tend to converge quicker than when errors are calculated using simulations from the same sample. We observe that error calculations are within 5\% of the converged value when considering 325 or more simulations (equivalent to the number of simulations in 25 samples).

Collectively, these results lend confidence in the quantification of global error via testing of random points in the parameter space considered. While the ternary plots we present in the previous section are useful in elucidating the distribution of error across the texture-weight space considered, we find that the results in Figure~\ref{fig:randerror} indicate that the estimation of a global mean error may be estimated using a significantly smaller number of simulations. This will become important in future studies where an increase in model complexity and dimensionality would require an intractably large number of simulations to estimate error as we have in the previously-discussed heat maps. Admittedly, however, we note that the methods in this section may not find the extreme errors, only the global mean.

\subsection{Higher-Dimensional Textures}
\label{subsec:highdimtex}

\begin{figure}
    \centering
    \subfigure[]{%
	\includegraphics[width=0.25\textwidth]{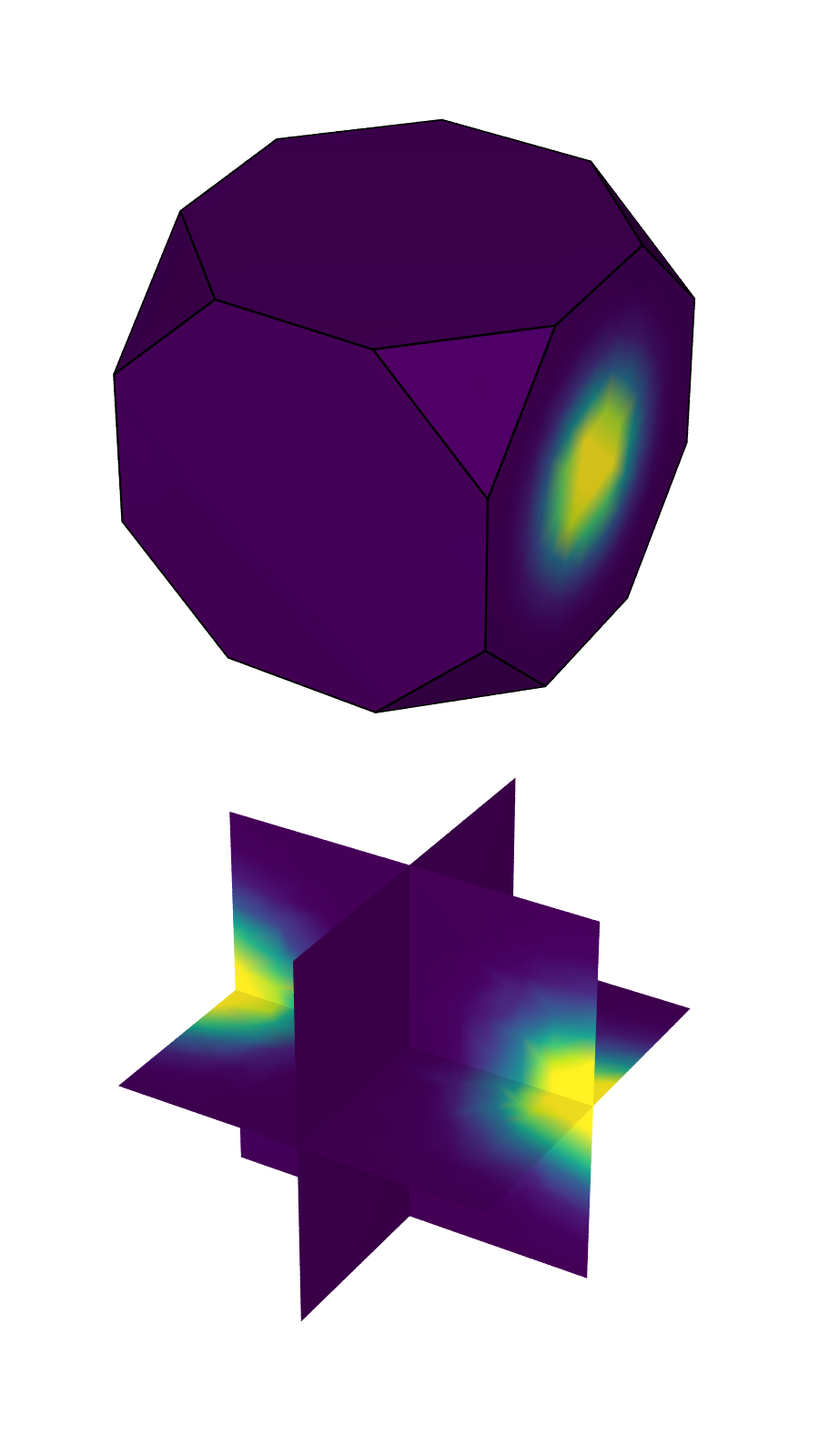}
        \label{subfig:newtex}}
    \subfigure[]{%
	\includegraphics[width=3.15in]{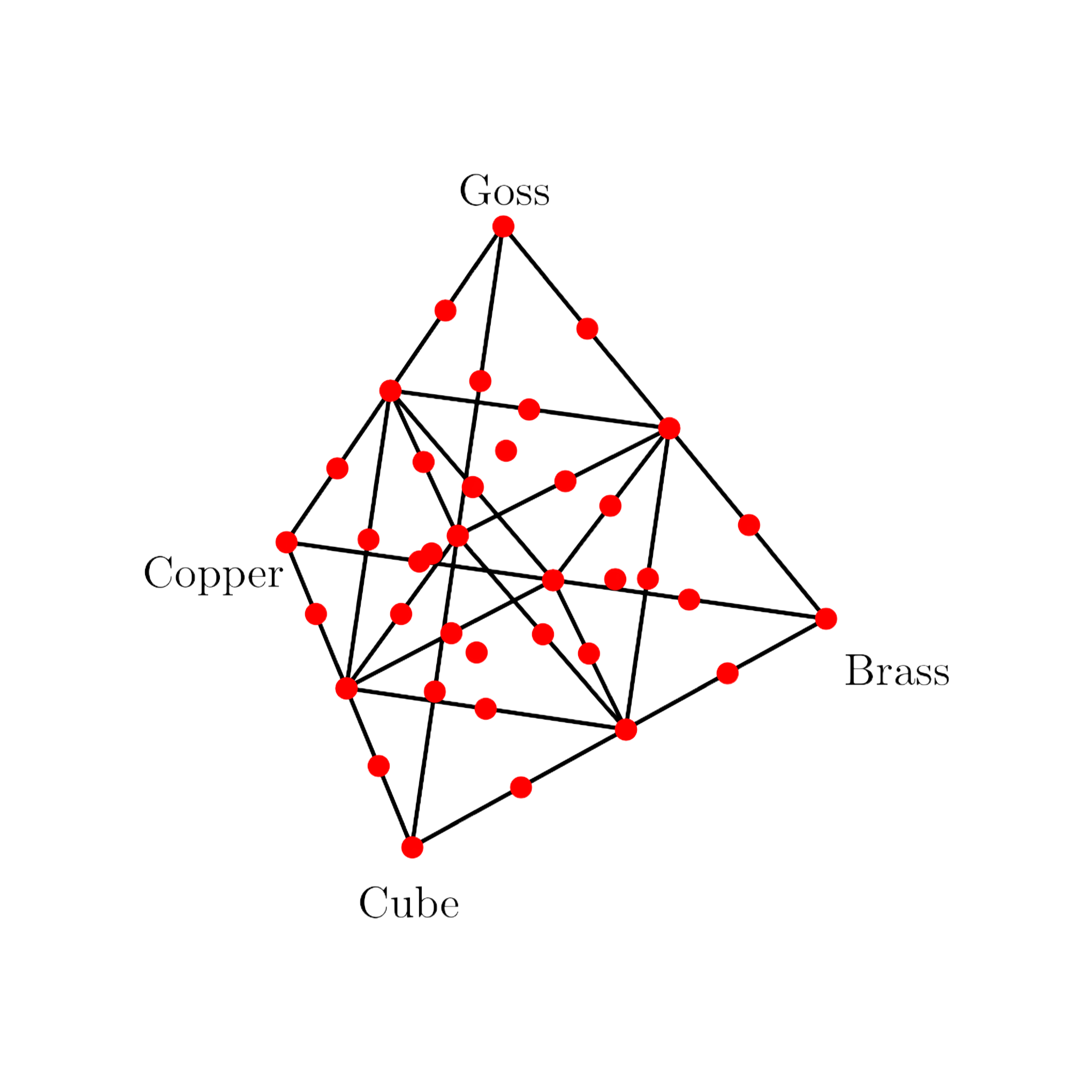}
        \label{subfig:quatdiagram}}
    \caption{~\subref{subfig:newtex} The Goss texture mode, considered in addition to the cube, brass, and copper texture modes, plotted on the same scale as in Figure~\ref{fig:extex}, and~\subref{subfig:randerrorsims} quaternary diagram depicting the texture weights of training points for samples with quadmodal textures (38 samples total), where the vertices represent unimodal texture distributions and the interior of the tetrahedron are points where the weights of all four textures sum to 1 (or 100\%).}
    \label{fig:quadmodal}
\end{figure}

\begin{figure}
    \centering
    \includegraphics[width=3.15in]{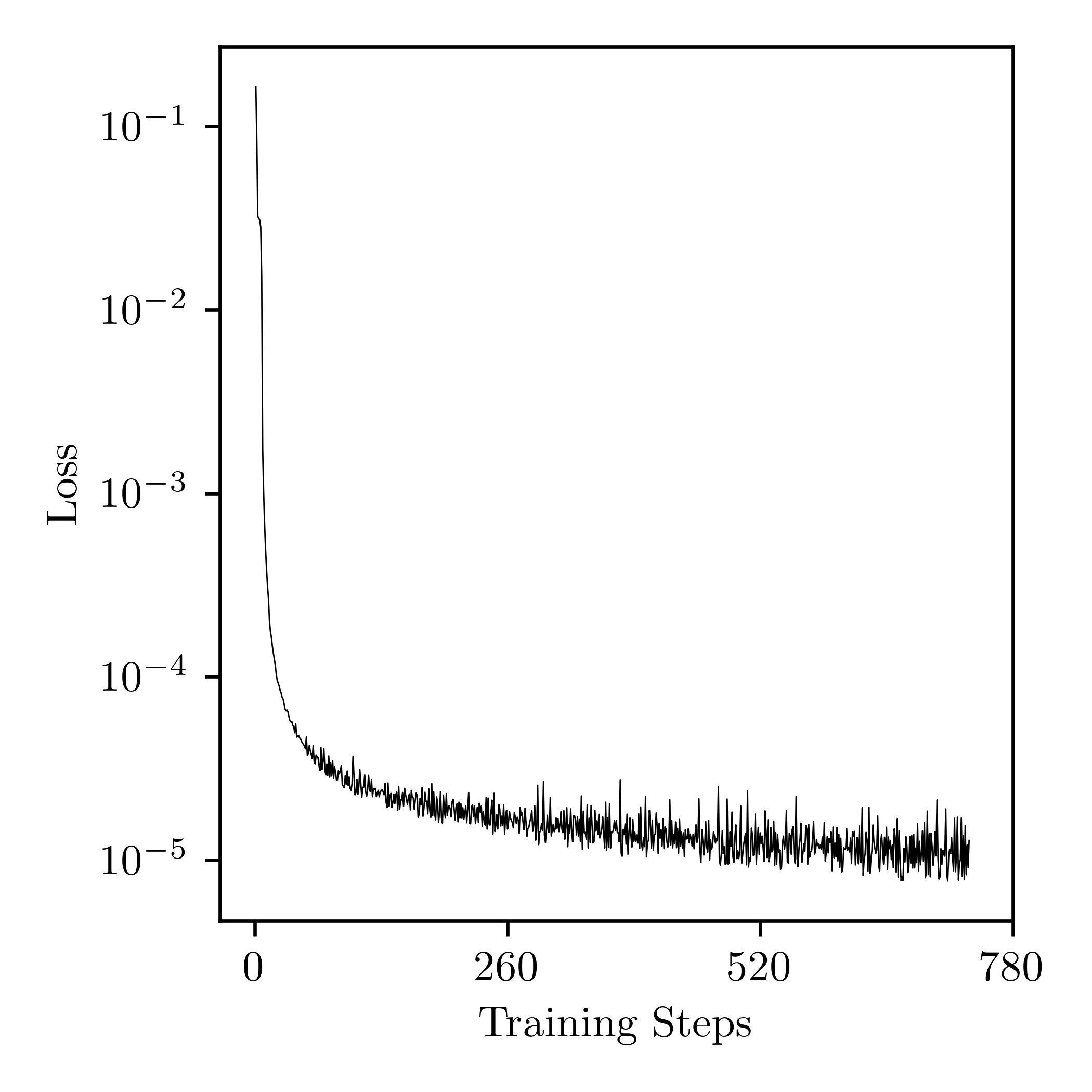}
    \caption{Training loss curve for the pICNN trained with simulations performed with quadmodal textures marked at the red points in Figure~\ref{subfig:quatdiagram} (i.e., 38 textures).}
    \label{fig:trainlossquad}
\end{figure}

In all cases above, we have considered at most three texture modes. Here, we explore the addition of a fourth texture mode in an effort to study the extensibility of our method considering high dimensional material description vectors. In particular, we wish to explore the consequences of the addition of more texture modes (which can broadly be seen as representative of \emph{any} extension to the material description vector). In particular, we consider the addition of the Goss texture mode, depicted in Figure~\ref{subfig:newtex}. We likewise consider here a similar recursive scheme as utilized for the ternary diagram~\citep{tetthesis} (see: Section~\ref{subsec:texture}) to generate a strategic set of training points in this higher dimensional space, which we plot for a common value of $\theta$ on a quaternary diagram in Figure~\ref{subfig:quatdiagram} (specifically using 38 samples/textures, or 494 total simulations, to train across this texture-weight space). We note that this quaternary diagram represents a plane in the four-dimensional texture-weight space. For the sake of computational expediency, we train to a lower number of training steps than previously, owing to an (expected) increase in training time as a function of the increase of the size of the training set and the model architecture. We present the training loss curve in Figure~\ref{fig:trainlossquad}. We note that we approach a very low loss (on the order of \SI{1e-5}{}), as well as saturation behavior, lending confidence in our decision to truncate training early compared to the trimodal models.

To gain as detailed a sense of the distribution of the MPE as for the trimodal samples, such a fine discretization of texture-weight space for the quadmodal samples would necessitate tens of thousands of simulations (as discussed in Section~\ref{subsec:randerror}). Consequently, we here utilize a random error calculation to efficiently estimate the global mean MPE across the texture-weight space considered. Consequently, we generate 1,000 simulations with random quadmodal textures and deformed along random load vectors (i.e., biaxialty ratios), and compare the predictions from the trained pICNN model to those from the CPFEM simulations. We find the mean error value to be 2.26\%, with a standard deviation of 2.03\%, and the maximum error to be 11.08\% (recall, this maximum error is for a single simulation, and may not be broadly representative of the entire yield surface for that texture, though we expect the mean and standard deviation to be representative of the global errors). Broadly, this indicates that the framework is adept at training models with increased dimensionality, lending further confidence in its applicability to generalized textures and increased material descriptor dimensionality (and possibly training via other texture representation means that require more descriptors, such as discrete spherical harmonics).

We further note that the trained quadmodal model generates a yield prediction in approximately \SI{5.02}{\second}, compared to \SI{2.60}{\second} for the trimodal model---i.e., a near two-fold increase. This indicates that the increase in complexity in the texture description has an impact on the prediction time of yield surfaces, again suggesting that a reduced-order texture descriptor is advantageous. The trained model is, of course, still orders of magnitude faster than the average time for a CPFEM simulation, or \SI{503}{\second} on average over 100 random CPFEM simulations performed with quadmodal texture distributions---a 100-fold reduction in time. We could likely employ recent developments in the generation of analytical expressions~\citep{fuhgextreme}, which could reduce the computational cost to the relatively cheap cost of the evaluation of an analytical expression.

%%%%%%%%%%%%%%%%%%%%%
\section{Conclusion and Future Directions}
\label{sec:Conclusion}

We have provided the motivation, formulation, and demonstration of a partially input convex neural network framework to train a model that predicts the macroscopic yield surface for polycrystalline samples with complex textures. Notably, the training is not performed on a case-by-case basis for each yield surface separately, but in a parametric way with respect to the texture. Via the generation of a relatively small set of strategic crystal plasticity finite element simulations, we were able to prove the framework's viability in relating complex material microstructures to the macroscopic behavior. This proves the extensibility of a previously developed framework first demonstrated on irreducibly-simplified material states to a significantly more complex problem. Specifically, we found that:
\begin{itemize}
    \item the framework is adept at generating a predictive model for yield surfaces of polycrystalline materials with complex crystallographic textures. We have also demonstrated the generalization of the framework through the consideration of increasingly higher-dimensional texture descriptions (to acceptable results),
    \item the inclusion of extra training points near transitional points of texture-weight space leads to overall better predictions and a reduction in peak error (indicating a non-linear relationship between texture modes and the resulting yield surface) while training the model with random points in parameter space generally leads to poor results, both results indicating the necessity for strategic placement of training points in parametric space, 
    \item the calculation of global error across the parameter space may be randomized and reduced for efficiency, and gives similar values as when considering a highly dense packing of simulations (which is inefficient as it requires a large amount of test data), 
    \item the model is adept at predicting yield surfaces of both highly textured materials, as well as those with more diffuse textures, lending confidence in the generality of the framework.
\end{itemize}

Overall, the results here lend us confidence that the framework demonstrated in this study will be adept at the inclusion of further material descriptors, including the consideration of more texture modes (as appropriate or physically necessary, or via the utilization of more complex texture reduction methods such as discrete spherical harmonics), more complex multiaxial loading states, material parameters, or geometric microstructural descriptors. We anticipate that the prime limitation of the development of an acceptable model is the generation of appropriately large data sets, though this study has demonstrated that strategic choices in the generation of training data allow for robust interpolative and extrapolative predictions by the model with relatively limited data. We likewise anticipate that this method can be used to generate predictions on limited sets of experimental data, which future work will explore. A fruitful exploration is also one in the direction of multifidelity learning, aiming to utilize experimental and synthetic data.

%%%%%%%%%%%%%%%%%%%%%
\section*{Acknowledgments}
\label{sec:acknowledgments}

The authors thank Dr. Romain Quey for the development of Neper. LvW and MK were funded or partially funded by Air Force Research Laboratory Grant FA8650-20-D-5211. KS was funded through the duration of this study by The Boeing Company as part of its Learning Together Program, and further received resources from the University of Alabama. PS and MO acknowledge funding from the Air Force Research Laboratory. JF and NB were supported by the SciAI Center, and funded by the Office of Naval Research (ONR), under Grant Number N00014-23-1-2729.

\bibliographystyle{unsrtnat}
\bibliography{references_arxiv}  %%% Uncomment this line and comment out the ``thebibliography'' section below to use the external .bib file (using bibtex) .





% \end{thebibliography}

\end{document}